\documentclass[journal]{vgtc}
\vgtccategory{Research}
\vgtcpapertype{Technique}

\usepackage{graphicx}%
\usepackage{booktabs}%
\usepackage{multirow}%
\usepackage{amsmath,amssymb,amsfonts}%
\usepackage{amsthm}%
\usepackage{mathrsfs}%
\usepackage[title]{appendix}%
\usepackage{textcomp}%
\usepackage{manyfoot}%
\usepackage{algorithm}%
\usepackage{algorithmicx}%
\usepackage{algpseudocode}%
\usepackage{listings}%

\usepackage{caption}
\usepackage{subcaption}
\usepackage{array}
\usepackage{amssymb}

\newcommand{\Skip}[1]{}

\newcommand{\DS}[1]{
   \textcolor{blue}{\bfseries{DS: {#1}}}
}

\newcommand{\shortcite}[1]{\cite{#1}}

\newcommand{\ToCheck}[1]{{{#1}}}

\begin{document}

\title{RT-HDIST: Ray-Tracing Core-based Hausdorff Distance Computation}

\author{%
  \authororcid{YoungWoo Kim}{0000-0003-3341-1714},
  \authororcid{Jaehong Lee}{0000-0002-8311-5975}, and
  \authororcid{Duksu Kim}{0000-0002-9075-3983}
}

\authorfooter{
  \item
  	YoungWoo Kim is with Korea University of Technology and Education(KOREATECH).
  	E-mail: aister9@koreatech.ac.kr.
  \item
  	Jaehong Lee is with Korea University of Technology and Education(KOREATECH).
  	E-mail: zerg100009@gmail.com.

  \item Duksu Kim is with Korea University of Technology and Education(KOREATECH).
  	E-mail: bluekdct@gmail.com.
}


\abstract{
The Hausdorff distance is a fundamental metric with widespread applications across various fields.  
However, its computation remains computationally expensive, especially for large-scale datasets.  
In this work, we present RT-HDIST, the first Hausdorff distance algorithm accelerated by ray-tracing cores (RT-cores).  
By reformulating the Hausdorff distance problem as a series of nearest-neighbor searches and introducing a novel quantized index space, RT-HDIST achieves significant reductions in computational overhead while maintaining exact results.  
Extensive benchmarks demonstrate up to a two-order-of-magnitude speedup over prior state-of-the-art methods, underscoring RT-HDIST's potential for real-time and large-scale applications.  
}

\keywords{Hausdorff distance, Ray-tracing core, Proximity query, General-purpose RT-core}

\teaser{
  \centering
  \includegraphics[width=\columnwidth]{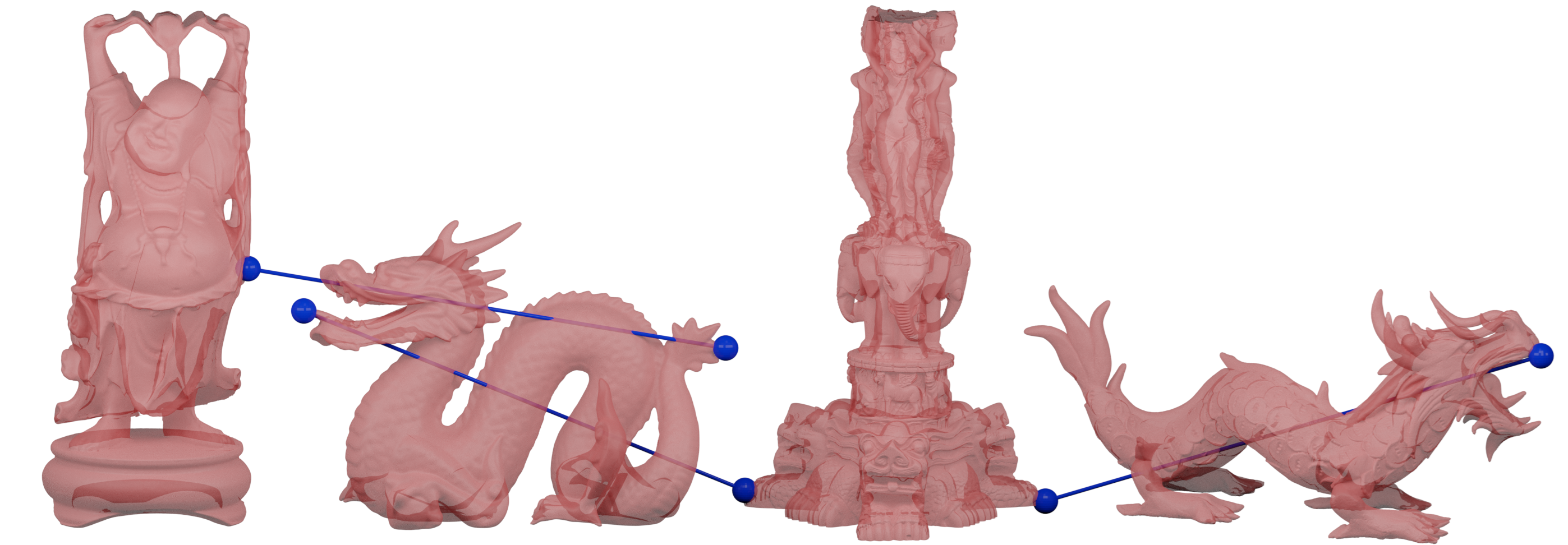}
  \caption{Benchmark models used in our experiment.  
    From left to right: Happy Buddha (543K), Dragon (437K), Thai Statue (4.9M), and Asian Dragon (3.6M).  
    The blue lines between objects represent the Hausdorff distance computed by our RT-HDIST algorithm.
  }
  \label{fig:teaser}
}

\date{April 2025}



\maketitle

\section{Introduction}

\Skip{
\begin{figure}
  \centering
  \includegraphics[width=\columnwidth]{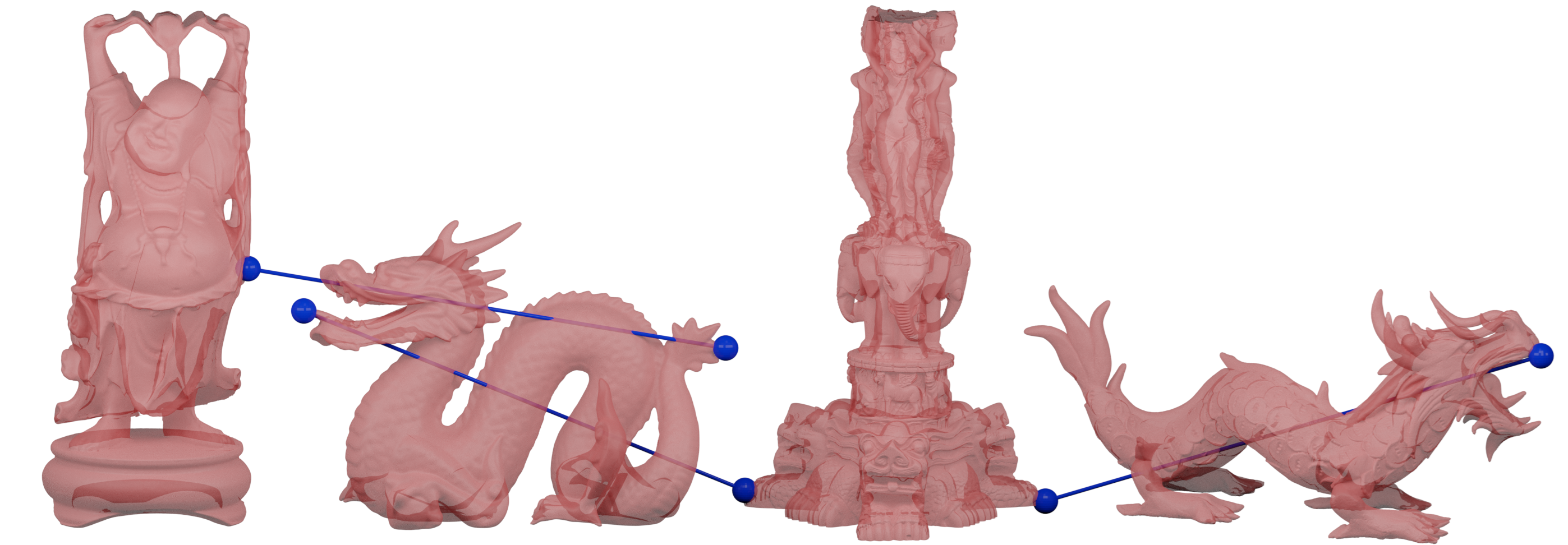}
  \caption{Benchmark models used in our experiment.  
    From left to right: Happy Buddha (543K), Dragon (437K), Thai Statue (4.9M), and Asian Dragon (3.6M).  
    The blue lines between objects represent the Hausdorff distance computed by our RT-HDIST algorithm.
  }
  \Description{This is the teaser figure for the article, showcasing four benchmark models and the performance of the RT-HDIST algorithm.}
  \label{fig:teaser}
\end{figure}
}

The Hausdorff distance is a fundamental metric for quantifying the similarity or dissimilarity between two geometric objects in a metric space. It finds extensive applications in fields such as computer graphics, computer vision, robotics, and more~\cite{tang2009interactive, trajectory2024}.  
However, computing the Hausdorff distance is computationally expensive, posing significant challenges when applied to large-scale datasets or real-time scenarios.


To address these challenges, researchers have developed various acceleration techniques, including hierarchical data structures like bounding volume hierarchies (BVH)~\cite{taha2015efficient, zheng2022economic}, culling-based methods~\cite{sacht2024cascading}, and approximation strategies such as sampling~\cite{ryu2021efficient}.  
While these methods have successfully reduced computation times, they often rely on CPU-based architectures and sequential processing, as they depend on processes like bounds updates, which require synchronization. This makes them less suitable for modern, massively parallel computing hardware such as GPUs.
Although some GPU-accelerated approaches have been explored for specific geometries such as freeform surfaces~\cite{krishnamurthy2011gpu}, a general-purpose solution for GPU-accelerated Hausdorff distance computation remains relatively unexplored.

In parallel, advancements in GPU hardware have introduced dedicated ray-tracing cores (RT-cores), originally developed to accelerate realistic rendering.  
Recent studies have explored the potential of RT-cores for non-rendering applications~\cite{wald2019rtx}, including nearest neighbor searches~\cite{zhu2022rtnn}, database indexing~\cite{henneberg2023rtindex}, and outlier detection~\cite{wang2024rtod}.  
These works highlight the versatility of RT-cores and their efficiency in handling spatial queries using BVHs.
Building on this momentum, we propose leveraging RT-cores to accelerate Hausdorff distance computation, presenting a novel approach that integrates RT-core hierarchical processing with GPU core geometric precision.

In this work, we introduce RT-HDIST, the first RT-core-based algorithm for Hausdorff distance computation.  
Our method reformulates the Hausdorff distance problem as a series of nearest-neighbor (NN) searches (Sec.~\ref{sec:NN-HDIST}).  
Building on this reformulation, we propose an RT-core-accelerated Hausdorff distance algorithm (Sec.~\ref{sec:RT-HDIST}) that efficiently leverages the hierarchical processing capabilities of RT-cores.  
To further optimize computation for large-scale problems, we introduce a novel quantized index space, which clusters target points into representative cells, significantly reducing computational overhead.  

We validate our method through extensive benchmarks conducted on three generations of GPUs equipped with RT-cores (Sec.~\ref{sec:results}).  
Our experimental results demonstrate that RT-HDIST achieves up to \ToCheck{195.24} times higher performance compared to prior state-of-the-art algorithm~\cite{sacht2024cascading}, while maintaining exact Hausdorff distance results.
These results demonstrate the efficiency of our method, making it suitable for use in interactive applications.



\section{Related Work}

\paragraph{Hausdorff Distance Computation}

The Hausdorff distance is a widely used query, applied in tasks such as measuring similarity between two subsets, penetration depth calculation, image analysis, and more~\cite{jesorsky2001robust,tang2009interactive,trajectory2024}.
To address its high computational cost, previous research has focused on developing faster methods, including sorting points~\cite{zhang2017efficient}, sampling strategies~\cite{ryu2021efficient}, and culling-based methods~\cite{taha2015efficient}.  

Among these, culling-based methods, which compute bounds for the Hausdorff distance and reduce the search space, have been most actively studied, as they can significantly reduce the computation time.
Tang et al.~\shortcite{tang2009interactive} proposed faster algorithms for triangle meshes by iteratively increasing lower bounds and decreasing upper bounds.
Zheng et al.~\shortcite{zheng2022economic} enhanced Tang's approach by introducing a four-point strategy, leveraging three triangle vertices and their center point.
Most recently, Sacht et al.~\shortcite{sacht2024cascading} proposed the Pompeiu-Hausdorff distance algorithm, which utilizes four upper bounds instead of one.
Their approach achieved up to 16 times faster performance compared to the Zheng et al.'s method~\shortcite{zheng2022economic}

Although culling approaches have achieved impressive performance improvements, they are approximate algorithms that operate within guaranteed error bounds (e.g., between lower and upper bounds).
These methods also face challenges in utilizing parallel hardware, as updating and sharing the bounds require synchronization, which is not well-suited for massively parallel processing architectures such as GPUs.
While a few studies have explored the use of GPUs for specific types of geometry, such as freeform surfaces~\cite{kim2010precise, krishnamurthy2011gpu}, employing GPUs for Hausdorff distance computation is relatively underexplored compared to other proximity computations~\cite{DSKim13, gProximity}.  

In this work, we propose an accurate Hausdorff distance algorithm accelerated by a GPU, specifically leveraging the ray-tracing core for enhanced performance.

\paragraph{General Purpose Ray Tracing Core}
Recent advancements in GPU technology have introduced dedicated ray-tracing cores, designed to enable hardware-accelerated ray tracing.
Originally developed for realistic rendering applications, RT-cores are now being explored for their potential to accelerate non-ray-tracing problems, much like GPU shader units, which have become widely used for general-purpose computing tasks.  

A pioneering work by Wald et al.~\shortcite{wald2019rtx} demonstrated the use of RT-cores to solve the tet-mesh point location problem.
Following this, RT-cores have been applied to a variety of problems.
For example, Thoman et al.~\shortcite{thoman2020rtx} adapted RT-cores for room response simulation and later extended their approach to support multiple GPUs~\cite{thoman2022multi}, achieving up to a 71x speedup using four GPUs compared to running on 12 CPU cores.  
Menesse et al.~\shortcite{meneses2024accelerating} implemented a range-minimum query algorithm using RT cores, achieving up to 5x and 2.3x faster performance compared to parallel CPU and GPU algorithms, respectively.
Wang et al.~\shortcite{wang2024rtod} introduced DODDS, an RT-core-based outlier detection algorithm, achieving up to 9.9x speedups over prior methods.
Zhu et al.~\shortcite{zhu2022rtnn} proposed the first nearest neighbor search using RT cores, which Nagarajan et al.~\shortcite{mandarapu2024arkade, nagarajan2023trueknn} extended, applying it to DBSCAN~\cite{nagarajan2023rtdbscan}.
Additionally, Henneberg et al.~\shortcite{henneberg2023rtindex} and Shi et al.~\shortcite{shi2024rtcudb} leveraged RT cores for database indexing and query systems, demonstrating their efficiency in non-ray-tracing applications.  

Building on this research trend, general-purpose RT-core utilization has emerged as a promising field. 
In this context, we propose the first RT-core-based algorithm for Hausdorff distance computation.  

\section{Preliminaries}

This section introduces the Hausdorff distance and RT-core technology.
Also, we introduce the RT-core-based $\epsilon$-neighbor search method from previous research, a basic component of our approach.

\subsection{Hausdorff Distance} \label{subsec:Hausdorff_distance}

The Hausdorff distance is formally defined in Eq.~\ref{equation:hausdorff_definition}.
Consider $A$ and $B$ as point clouds (e.g., sets of vertices in an object), and let $d(\cdot, \cdot)$ denote the Euclidean distance between any two points.

\begin{equation}
    H(A,B) = \max \left( \max_{a \in A} \min_{b \in B} d(a,b) ,
    \max_{b \in B} \min_{a \in A} d(b,a) \right)
    \label{equation:hausdorff_definition}
\end{equation}

Additionally, the one-sided Hausdorff distance can be defined as shown in Eq.~\ref{equation:oneside_hausdorff}. The Hausdorff distance, $H(A,B)$, is then represented as the maximum of the two one-sided distances: $H(A,B) = \max(h(A,B), h(B,A))$.

\begin{equation}
    h(A,B) = \max_{a \in A} \min_{b \in B} d(a,b)
    \label{equation:oneside_hausdorff}
\end{equation}

The brute-force computation of the Hausdorff distance has a time complexity of $O(nm)$, where $n$ and $m$ represent the number of points in the two point clouds, as it involves testing all pairs of points. To address this computational overhead by reducing the search space, acceleration hierarchies like Bounding Volume Hierarchies (BVH) are widely employed.

Recent research has emphasized improving the traversal efficiency of such hierarchies by rapidly narrowing the upper and lower bounds of the target value and utilizing comparisons between point-to-bounding-volume (BV) distances to streamline the computation~\cite{zheng2022economic, sacht2024cascading}.
However, these state-of-the-art Hausdorff distance algorithms are typically implemented on CPUs because updating and sharing the bounds require synchronization, which is not well-suited for massively parallel processing.  


\subsection{Ray-Tracing Core}\label{subsec:RT-core}

The RT-core is a specialized hardware component designed to accelerate ray-tracing-based rendering.
It was first introduced by NVIDIA as part of the RTX platform~\cite{RTXon}, and more recently, many GPUs have incorporated similar ray-tracing hardware~\cite{amd_rdna2_2020,intel_raytracing_guide}.
At the core of ray tracing lies the essential operation of intersection testing between rays and objects.
The RT-core significantly enhances this process by offering hardware-based capabilities for efficient ray-BV and ray-polygon (e.g., triangle) intersection tests.

To fully utilize the capabilities of this specialized hardware, the target task must be structured as a series of intersection tests between rays and BV or polygons.
Although distance computations, including Hausdorff distance, are typically accelerated using an acceleration hierarchy (e.g., BVH) based on the distance between a query point and a bounding volume~\cite{Larsen00, zheng2022economic}, the RT-core platform does not support point-BV distance calculations directly.

To address this limitation in the context of accelerating Hausdorff distance computation using the RT-core, we redesign the broad-phase computations to rely on a series of ray-BV tests.

\subsection{RT-based $\epsilon$-Neighbor Search}

Recent research has effectively utilized RT-cores for accelerating distance computations, particularly in k-nearest neighbor searches~\cite{zhu2022rtnn, nagarajan2023trueknn,mandarapu2024arkade}.
This is typically achieved through a two-phase approach: a broad phase, leveraging the RT-core to identify candidate points, followed by a narrow phase, performing precise computations on GPU cores (e.g., CUDA cores).

The broad phase employs an RT-core-based $\epsilon$-neighbor search to identify points within a radius $\epsilon$ of a query point $q$.
Given a target point cloud $P = \{p_1, p_2, ..., p_n\}$, an axis-aligned bounding box (AABB), denoted as $BV(p_i, \epsilon)$, is constructed for each point $p_i \in P$, centered at $p_i$ with side lengths $2\epsilon$.
Current RT-core platforms do not support intersection test with spherical volumes, necessitating the use of AABBs.
A BVH is then built over these AABBs.
For each query point $q$, a short ray $ray(q)$ is cast from $q$ in an arbitrary direction.

The RT-core efficiently identifies all AABBs that intersect with $ray(q)$.
The points $p_i$ corresponding to these intersecting AABBs constitute the candidate set of $\epsilon$-neighbors of $q$:

\begin{equation}
N(q, \epsilon) = \{ p_i \in P \mid BV(p_i, \epsilon) \cap ray(q) \neq \emptyset \}
\end{equation}

Once the broad phase identifies candidate points, the narrow phase employs precise distance computations on standard GPU cores to confirm which points truly lie within \(\epsilon\) of \(q\) or to sort them by their actual distance.
This approach constitutes a building block of our proposed RT-HDIST algorithm.

\section{NN-based Hausdorff distance algorithm}\label{sec:NN-HDIST}

To accelerate Hausdorff distance computation using RT-core, we reformulate the problem as a series of nearest-neighbor (NN) searches.
This section presents the NN-based algorithm, foundational for its RT-core implementation (Sec.~\ref{sec:RT-HDIST}).

\subsection{Problem Formulation}
Assuming spherical volumes for the target points, if \(N(a_i, r_k)\) is non-empty for some \(r_k\), the nearest neighbor distance of \(a_i\) relative to \(B\) can be defined as  
\begin{equation}
   N(a_i) \;=\; \min_{b_j \in N(a_i, r_k)} d(a_i, b_j),    
\end{equation}
where \(a_i \in A\) and \(b_j \in B\).
Consequently, the one-sided Hausdorff distance is
\begin{equation}
   h(A, B) \;=\; \max_{a_i \in A} N(a_i).
\end{equation}

To bracket \(h(A,B)\), note that if there exists some \(a_i\) uncovered by \(r_k\) (i.e., \(N(a_i, r_k) = \emptyset\)), then \(\min_{b_j\in B}d(a_i,b_j)>r_k\). Hence
\begin{equation}
   h(A,B) \;>\; r_k.
\label{eq:lower_bound}
\end{equation}
On the other hand, if a larger radius \(r_{k+1}\) covers every \(a_i\) (meaning \(N(a_i, r_{k+1})\neq \emptyset\) for all \(i\)), then 
\begin{equation}
   h(A,B)\;\le\;r_{k+1}.
\label{eq:upper_bound}
\end{equation}
Thus,
\begin{equation}
   r_k \;<\; h(A,B)\;\le\;r_{k+1}.
\end{equation}

Because only newly included neighbors in \(\,N(a_i, r_{k+1}) \setminus N(a_i, r_k)\)  
can push the distance into \((r_k,\,r_{k+1}]\), these form the candidate set for identifying \(h(A,B)\).
In other words, if \(r_k\) is the largest radius that still fails to cover some \(a_i\), then the exact Hausdorff distance must lie in the interval \((r_k,r_{k+1}]\).

\subsection{NN-HDIST Algorithm}
Based on our problem formulation, the broad phase of the NN-based Hausdorff distance (NN-HDIST) algorithm iteratively updates candidate sets.  

Starting with an initial radius \(r_0\), we first compute \(N(a_i, r_0)\) for every \(a_i \in A\).  
If \(N(a_i, r_0)\) is non-empty, the nearest neighbor of \(a_i\) lies within \(N(a_i, r_0)\).  
Hence, we record the pair \(C_{a_i} = \bigl(a_i, N(a_i, r_0)\bigr)\) as a candidate for \(h(A, B)\) and remove \(a_i\) from \(A\).  
If any points remain in \(A\) at this stage (i.e., \(A \neq \emptyset\)), we discard these candidate sets, since at least one point is still uncovered (Eq.~\ref{eq:lower_bound}).
Next, we increment the radius by \(\alpha\), giving a new radius \(r=r_0 + i \cdot \alpha\) (where \(i\) is the iteration count).  
We then repeat the neighbor search \(N(a_i, r)\) for the remaining query points in \(A\).  
Again, any newly covered points contribute to the candidate sets \(C_{a_i}\) and are removed from \(A\).  
This process continues until all query points are covered (Eq.~\ref{eq:upper_bound}).  
At the final iteration, the candidate sets \(C_{a_i}\) form the complete set of candidate pairs for \(h(A, B)\).  
Algorithm~\ref{alg:nn_hdist} provides the corresponding pseudo-code.

\begin{algorithm}[t]
\caption{Broad phase of NN-HDIST (one-sided: $h(A,B)$)}
\label{alg:nn_hdist}
\begin{algorithmic}[1]
\Function{BroadPhase}{$A, B, r_0, \alpha$}
  \State $r \gets r_0$

  \While{$A \neq \emptyset$}
    \State $C \gets \emptyset$ \quad // Initialize candidate set
    \ForAll{$a_i \in A$}
      \If{$N(a_i,\,r) \neq \emptyset$}
        \State $C_{a_i} \gets (a_i,N(a_i,r))$
        \State $C \gets C \cup C_{a_i}$
        \State $A \gets A \setminus \{a_i\}$
      \EndIf
    \EndFor
    \State $r \gets r + \alpha$
  \EndWhile
  \State \Return $C$
\EndFunction
\end{algorithmic}
\end{algorithm}

In the narrow phase, for each final candidate set $C_{a_i}$, it computes \(\max_{a_i \in A} \min_{b_j \in C_{a_i}} d(a_i, b_j)\), thus determining \(h(A, B)\).
By repeating this process for both sides, the final Hausdorff distance \(H(A, B)\) is determined.
\section{RT-HDIST algorithm}\label{sec:RT-HDIST}

In this section, we introduce our RT-based Hausdorff distance algorithm (RT-HDIST), which adapts the NN-HDIST algorithm to efficiently leverage RT-cores.

\subsection{Overview}
\begin{figure*}[t]
    \centering
    \includegraphics[width=\linewidth]{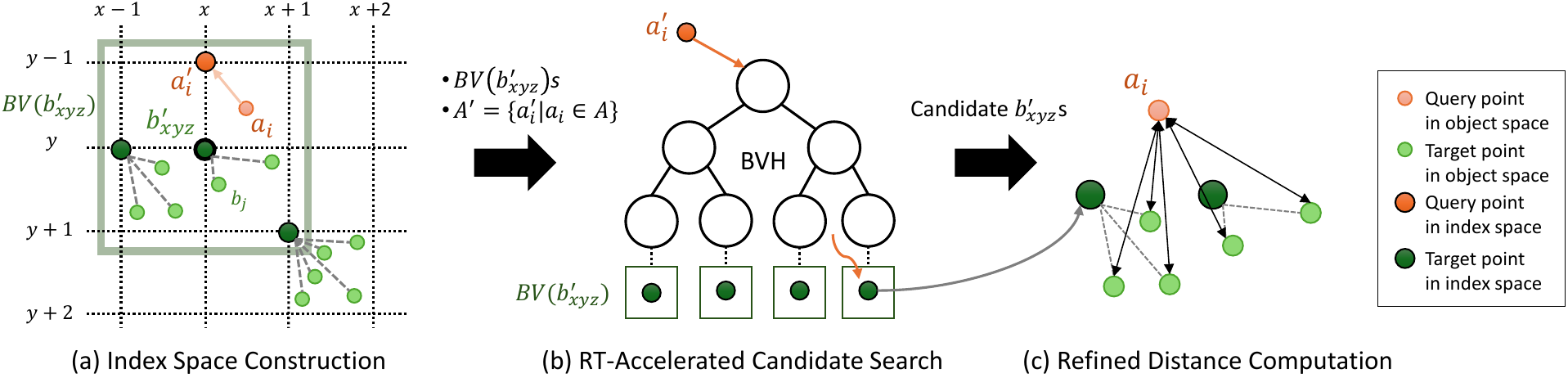}
    \caption{Overview of RT-HDIST algorithm
    }
    \label{fig:Overview}
\end{figure*}

A naive approach to implement the NN-HDIST algorithm on an RT-core platform involves setting an AABB for all target points (e.g., \(b_j \in B\)), building a BVH from them, and casting rays from every query point (e.g., \(a_i \in A\)).
However, we found that this approach is time-consuming and, for large-scale problems with many points, it can even be slower than prior CPU-based methods.

To address the challenges posed by large problem sizes, we employ a quantized space, referred to as the index space.
Fig.~\ref{fig:Overview} illustrates the overall process of our RT-HDIST algorithm.

In the \textbf{Index Space Construction} step (Sec.~\ref{subsec:index_space}), target points from the object space are projected into a quantized index space, resulting in points in the index space (e.g., \(b'_{xyz}\)).
We define \(b'_{xyz}\) as the \textit{representative point} located at \((x, y, z) \in \mathbb{Z}^3\) in the index space, while maintaining a list of points mapped to each representative point.
For every \(b'_{xyz}\), an AABB (\(BV(b'_{xyz})\)) is prepared.
Additionally, a query point set is generated in the index space (\(A'\)).
This step is executed on general GPU cores (e.g., CUDA cores), and the resulting \(BV(b'_{xyz})\)s and \(A'\) are passed to the next step.

The \textbf{RT-Accelerated Candidate Search} step (Sec.~\ref{subsec:RT-search}) operates entirely in the index space, leveraging the RT-core for efficient computation.
A BVH is constructed over the \(BV(b'_{xyz})\)s, with the specific BVH implementation depending on the RT-core platform (e.g., Geometry Acceleration Structure for NVIDIA RTX GPUs).
Once the BVH is constructed, the NN-HDIST algorithm (Sec.~\ref{sec:NN-HDIST}) is executed on the RT-core to perform the candidate search.

Finally, in the \textbf{Refined Distance Computation} step (Fig.~\ref{fig:Overview}-(c)), the candidate sets obtained from the prior step are mapped back to the object space.
This step is executed on general GPU cores to finalize the Hausdorff distance computation.

\subsection{Index Space Construction}
\label{subsec:index_space}

When a grid is defined, the index space consists of points that are located only at the corners of the grid cells.
In other words, the index of a cell corresponds to a point in the index space.
To differentiate points in the index space from those in the object space, we use the notation $'$ (prime) to denote points in the index space (e.g., $p')$.
Fig.~\ref{fig:Overview}-(a) illustrates an example of the index space, where the light-colored points represent points in the object space, while the dark-colored points with black outlines represent points in the index points.

For a given target point cloud $B$, a grid is constructed to define the index space.
This grid is based on the AABB that encloses $B$ (i.e., $BV(B)$).
The grid is composed of cubic cells, with the number of cells along each axis of $BV(B)$ determined by the grid resolution.
Let $l$ represent the length of the longest axis of $BV(B)$.
This axis is divided into $2^k$ cells, where $k \in \mathbb{Z}$ is a parameter referred to as the \textit{bit count}.
The size of each cubic cell is given by $s = l / 2^k$.
Consequently, the number of cells along each dimension of $BV(B)$ is $l_{\text{dim}} / s$, where $l_{\text{dim}}$ is the length of the dimension being considered.

\paragraph{Mapping Points to Index Space}  
Once the index space is prepared, the target points (e.g., $b_j$) are projected into the index space.  
To reduce the computational burden and improve efficiency, all target points located within the same cell $(x, y, z)$ are grouped together and represented as a single \textit{representative point} $b'_{xyz}$ in the index space.  
Geometrically, $b'_{xyz}$ is located at the corner of the cell with the minimum $x$, $y$, and $z$ coordinates.
The set of all representative points forms the target point cloud in the index space, $B' = \{b'_{xyz} \mid \text{cell}(x, y, z) \neq \emptyset\}$.

Similarly, the query points are projected into the index space.
Unlike the target points, query points within the same cell remain as individual points, as each query point serves as a basic unit for parallel processing in subsequent steps.
This process results in the query point set in the index space, $A' = \{a_i' \mid a_i \in A\}$.  
Notably, even if a query point lies outside the BV of the target point cloud, its corresponding index can still be determined using the cell size.

\paragraph{Defining BVs for Representative Points}  
To leverage the RT-core platform effectively, the RT-HDIST algorithm adopts AABBs as bounding volumes (BVs).
A point itself does not inherently have a BV, so it is necessary to define a BV for constructing the BVH.  
To ensure the initial volume covers all neighboring points in the index space, the BV for a representative point is defined as a $2\sqrt{3} \times 2\sqrt{3} \times 2\sqrt{3}$ AABB centered on the representative point.  

\paragraph{Bit Count Selection}  
The bit count determines the clustering level, which impacts the number of points within each cell and plays a crucial role in the overall performance of our method.
A higher clustering level results in faster performance during the RT-accelerated candidate search step because fewer representative points need to be processed.
However, it can lead to an increased number of final candidates in the refined distance computation step, potentially offsetting the performance gains.  
On the other hand, a lower clustering level reduces the number of final candidates but increases the computational load during the RT-accelerated candidate search step due to a higher number of representative points.

Although the optimal bit count may vary depending on the scene, our experiments indicate that a bit count between 6 and 8 generally yields good performance across various cases.

\subsection{RT-Accelerated Candidate Search}
\label{subsec:RT-search}

This step identifies candidate points for Hausdorff distance computation.
Handling all points in the object space introduces considerable computational and spatial overhead, stemming from the large BVH construction and lengthy traversal times.
Although using intermediate BVH nodes may appear to be a solution, current RT-core platforms do not support access to intermediate nodes. 
To address these limitations, the search is performed in the index space, effectively reducing overhead and localizing the search area.  

In this step, the RT-core executes the broad phase of the NN-HDIST algorithm.  
The initial volume size is determined by $r_0$, and the increment is set to the same value to ensure that the search region expands by at least one additional cell in all directions.  
In summary, the process performs $BroadPhase(A', B', \sqrt{3}, \sqrt{3})$.  
Since each query point is represented as a ray, query points are processed in parallel by the RT-cores, enabling efficient and scalable computation. 

\paragraph{Resolving AABB-Sphere Discrepancy}
One key difference between the NN-HDIST algorithm and RT-HDIST is the use of AABBs in RT-HDIST, whereas NN-HDIST assumes spherical volumes.
This difference can introduce errors in the $\epsilon$-neighbor search (e.g., $N(a'_i, r)$), as the AABB for a target point farther than $r$ may still overlap with the query point.  

To address this error, we perform an in-sphere test for $a'_i$s that overlap with the AABB to ensure accurate candidate selection, where $d(b'_{xyz}, a'_i) \leq r$.
The in-sphere test is conducted on the GPU core, while the AABB test is handled by the RT-core.  
Since the GPU cores are utilized, we can refine candidate query points more precisely by using the original query point $a_i$ instead of $a'_i$.
This is accomplished by scaling the coordinate values of $a_i$ to align with the index space.

\paragraph{Handling Quantization Error}
\begin{figure}[t]
    \centering
    \includegraphics[width=0.9\linewidth]{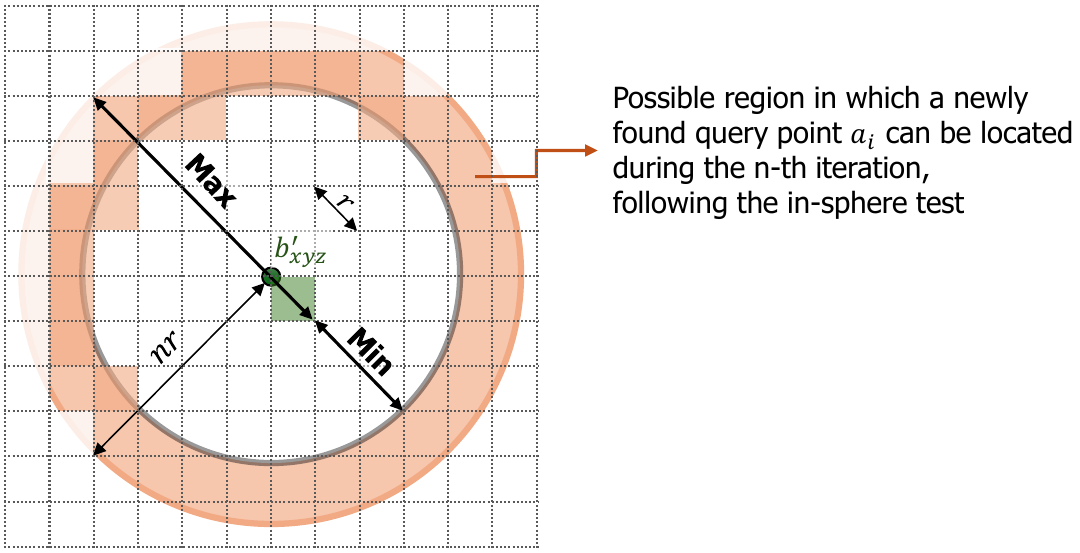}
    \caption{
    This figure illustrates the lower and upper bounds of $d(a_i, b_j)$, where $b_j$ is a target point affiliated with $b'_{xyz}$, and $a_i$ is a query point associated with $a'_i$, which was first identified at the $n$-th iteration.
    }
    \label{fig:bounds}
\end{figure}

The index space projection itself can introduce errors due to the quantization of points into the index space, as a target point $b_j$ affiliated with $b'_{xyz}$ could be located anywhere within the cell (green region in Fig.~\ref{fig:bounds}).
Fig.~\ref{fig:bounds} illustrates the upper and lower bounds of $d(a_i, b_j)$ in the index space, where $a_i$ is a query point first identified during the $n$-th iteration.

When $r_0 = \alpha$, both denoted as $r$, the lower bound of $d(a_i, b_j)$ is $(n-2)r$, which occurs when $b_j$ is located at the corner of the cell and $a_i$ is positioned at the closest diagonal point within the cell.  
Conversely, the upper bound occurs when $a_i$ is positioned at the farthest point along the opposite corner from $b_j$.
This distance cannot be larger than $(n+1)r$ because a sphere with radius $(n+1)r$ centered at any corner of the cell associated with $b'_{xyz}$ will enclose a sphere with radius $nr$ centered at $b'_{xyz}$, provided $r \geq \sqrt{3}$.  
Note that the center of the sphere can shift by up to $\sqrt{3}$ within a unit cube (edge length of 1).
As a result, the bounds of the actual distance $d(a_i, b_j)$ at the $n$-th iteration are as follows:  
\begin{equation}
    (n-2)r < N(a_i, nr) < (n+1)r
\end{equation}  

This implies that the candidates acquired at the $n$-th iteration are always greater than those from the $(n-3)$-th iteration and, therefore, do not contribute to the Hausdorff distance, which is determined by the maximum distance pair.  
In other words, to ensure accurate results, it is necessary to maintain the candidates from the current iteration and the two preceding iterations, a range referred to as the gray zone.  

To incorporate the error bounds into Algorithm~\ref{alg:nn_hdist}, we modify the algorithm to maintain the query points and their related candidate pairs as long as they remain in the gray zone.
Once a query point exits the gray zone, the point and its associated candidate pairs are discarded. 
Finally, the algorithm returns all remaining candidate pairs at the final iteration.

\section{Results and Analysis}\label{sec:results}

We implemented our RT-HDIST algorithm (i.e., \textit{Ours}) in C++, CUDA 12.3, and OptiX 7.4.  
The initial search radius (i.e., $r_0$) was set to the nearest $nr$ value corresponding to half the distance between the centers of the two objects.
All experiments were conducted on the same host environment, which includes an Intel i5-14600K CPU and 32GB of RAM.  
To evaluate the performance of our algorithm, we tested it on three different device environments, as detailed in Table~\ref{table:GPUs}, each representing a different generation of the RT-core platform.  

\begin{table}[t]
\centering
\caption{Specifications of GPUs used in the experiments}
\label{table:GPUs}
\small
\begin{tabular}{|c|c|c|c|}
\hline
{GPU}                  & {RTX2080} & {RTX3080} & {RTX4080}      \\ \hline
{Architecture}         & Turing           & Ampere           & Ada Lovelace          \\ \hline
{\# of CUDA cores}     & 2,944            & 8,704            & 9,728                 \\ \hline
{CUDA core clock}      & 1515 MHz         & 1450 MHz         & 2205 MHz              \\ \hline
{\# of RT cores (Gen.)}       & 46 (1st)               & 68 (2nd)               & 76 (3rd)                    \\ \hline
\end{tabular}
\end{table}

To evaluate the benefits of our method, we compared its performance with two prior state-of-the-art Hausdorff distance algorithms running on CPU, using their source code provided on GitHub:  
\begin{itemize}  
    \item \textit{Zheng}: This is an implementation of Zheng et al.'s~\shortcite{zheng2022economic} algorithm, with the stop condition $U - L < L * e$, where $U$ and $L$ represent the upper and lower bounds, respectively, and $e$ is set to 0.01.
    \item \textit{Sacht}: This is an implementation of Sacht and Alec's~\shortcite{sacht2024cascading} method, configured with an error tolerance of $\epsilon = 10^{-8}$ and a maximum subdivision factor $f = 10^{6}$.  
\end{itemize}

\begin{figure*}[h]
    \centering
    \begin{subfigure}{0.23\linewidth}
        \includegraphics[width=\textwidth]{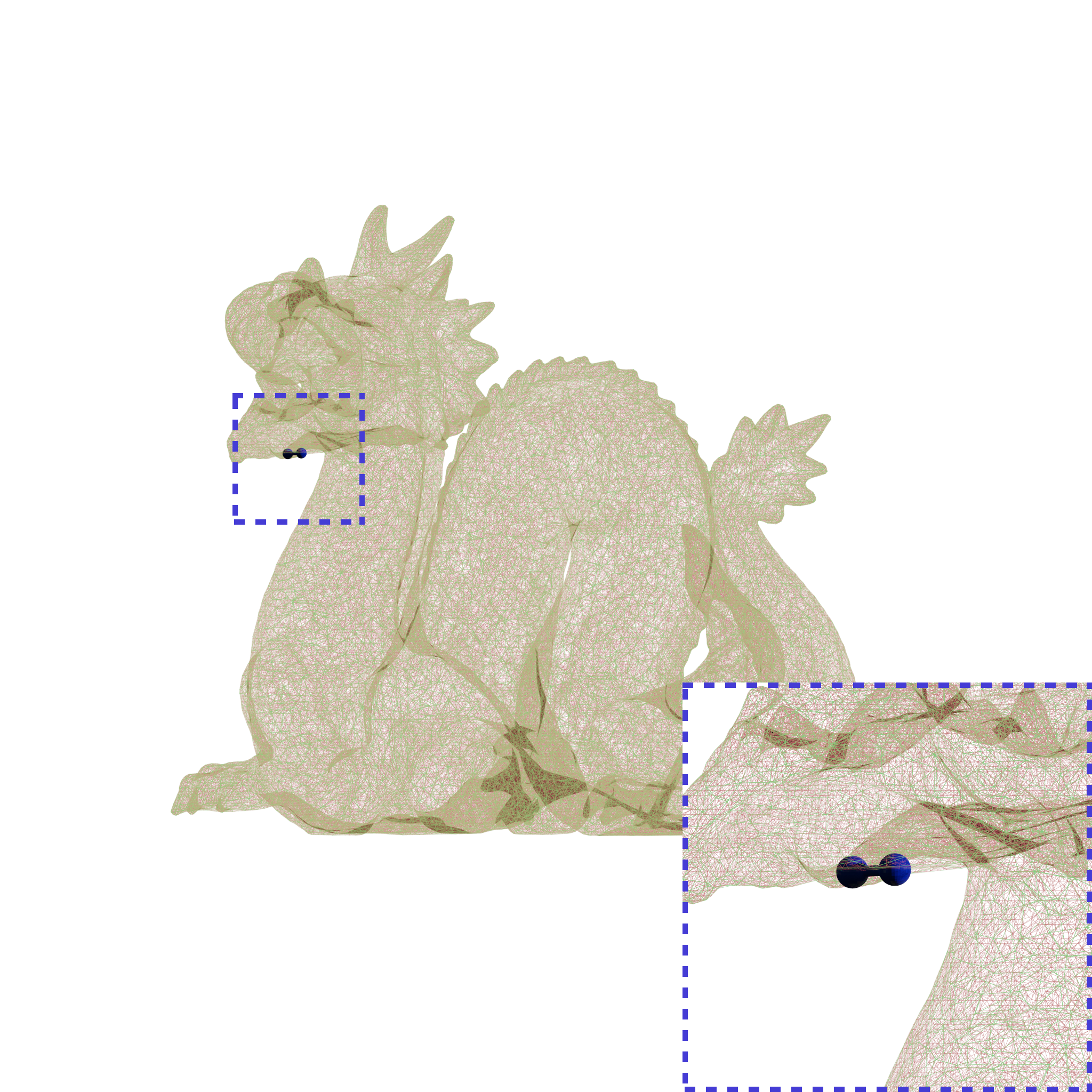}
        \caption{Dragon}
    \end{subfigure}
    \begin{subfigure}{0.23\linewidth}
        \includegraphics[width=\textwidth]{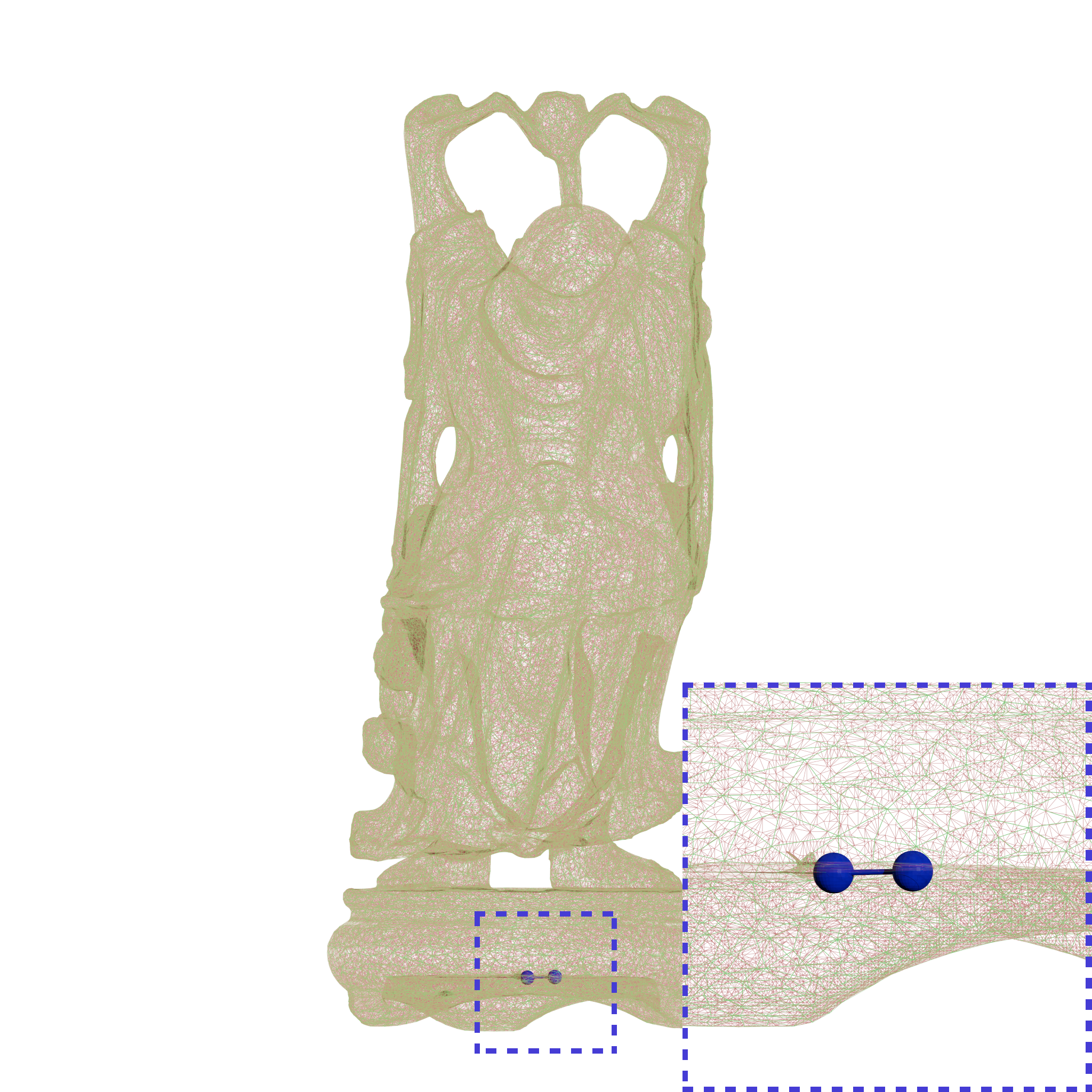}
        \caption{Happy Buddha}
    \end{subfigure}
    \begin{subfigure}{0.23\linewidth}
        \includegraphics[width=\textwidth]{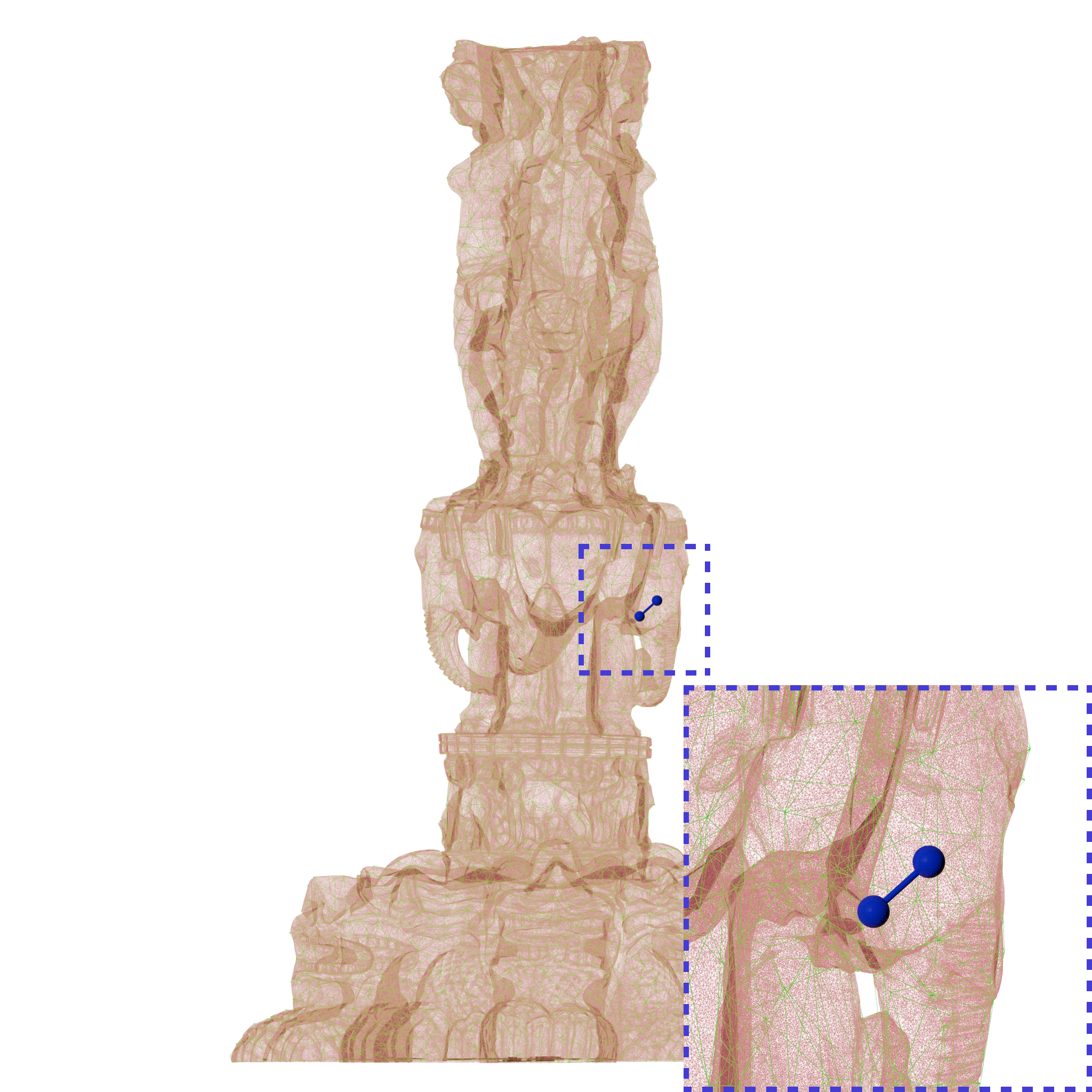}
        \caption{Thai Statue}
    \end{subfigure}
    \begin{subfigure}{0.23\linewidth}
        \includegraphics[width=\textwidth]{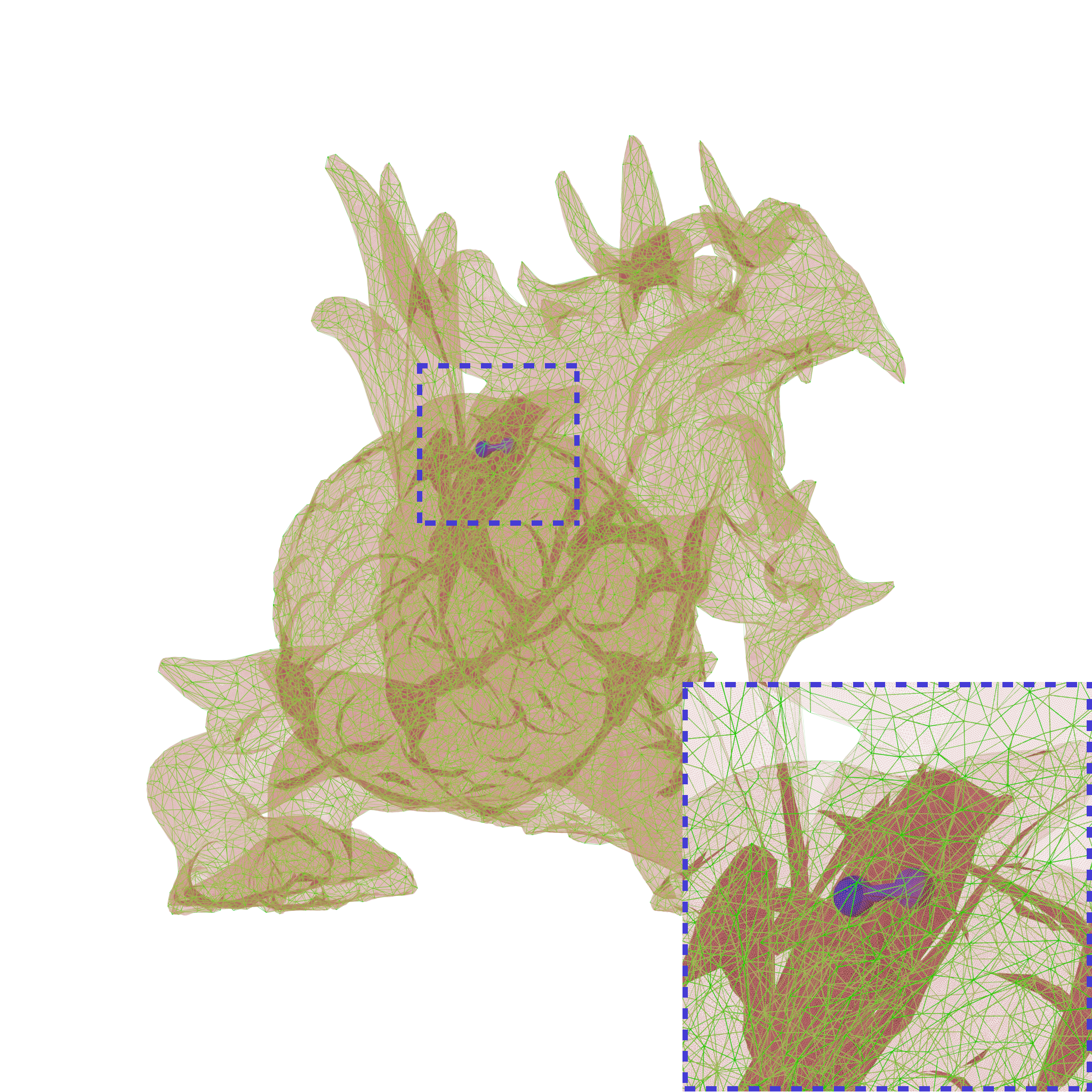}
        \caption{Asian Dragon}
    \end{subfigure}
    
    \caption{Results of the decimation benchmarks. The red mesh represents the original model, while the green mesh corresponds to the decimated version.}
    \label{fig:decimate_result}
\end{figure*}

\begin{figure*}[h]
    \centering
    \begin{subfigure}{0.23\linewidth}
        \includegraphics[width=\textwidth]{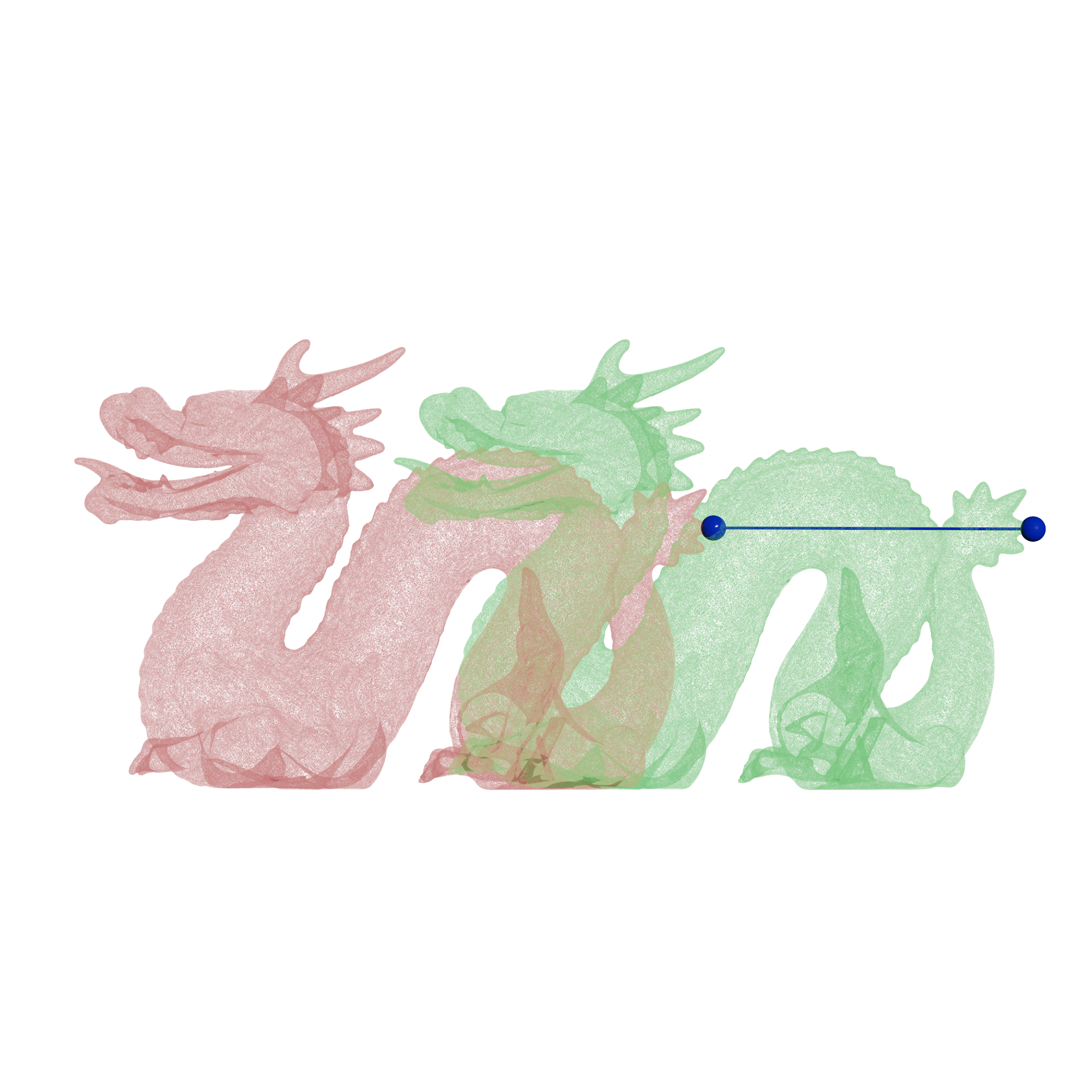}
        \caption{Dragon}
    \end{subfigure}
    \begin{subfigure}{0.23\linewidth}
        \includegraphics[width=\textwidth]{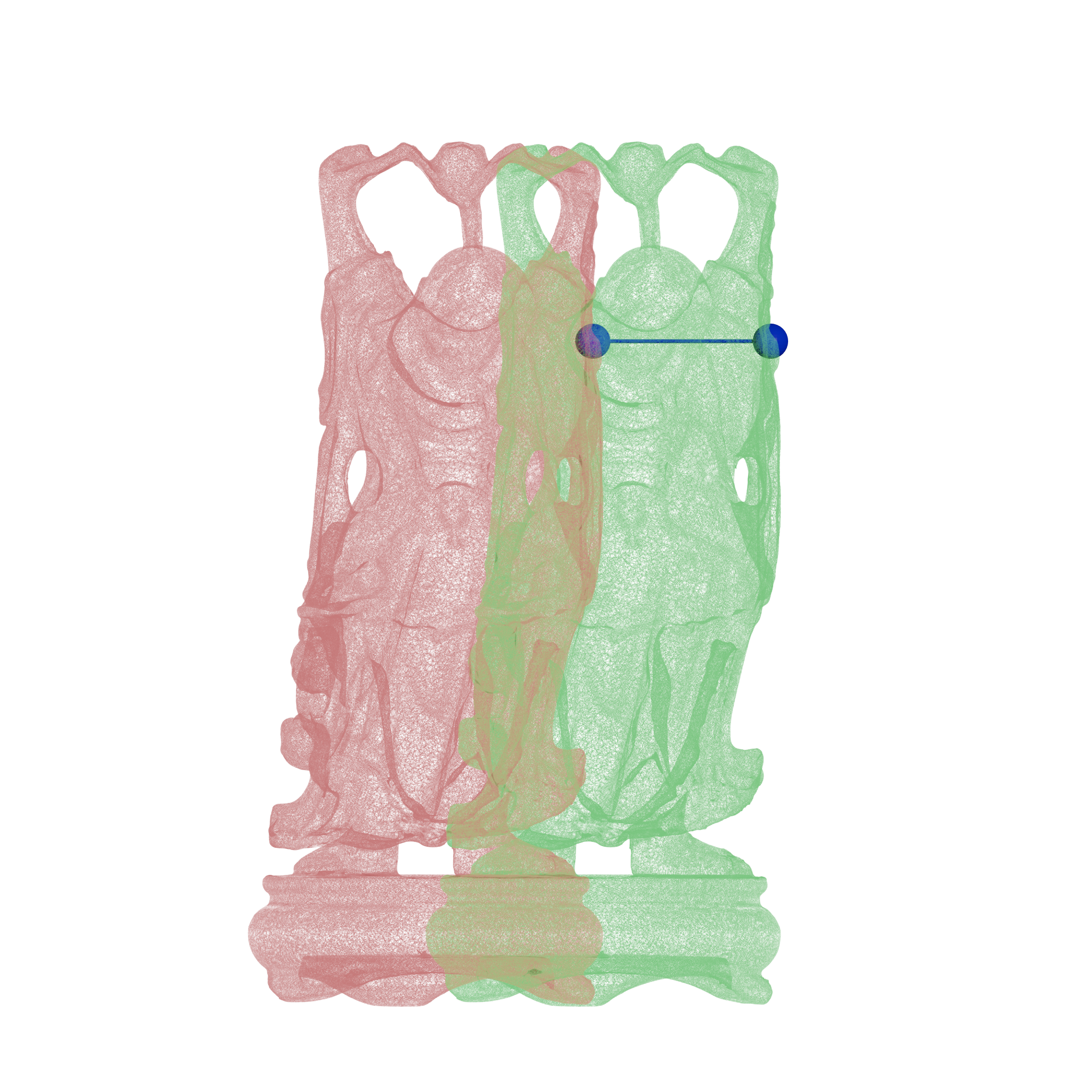}
        \caption{Happy Buddha}
    \end{subfigure}
    \begin{subfigure}{0.23\linewidth}
        \includegraphics[width=\textwidth]{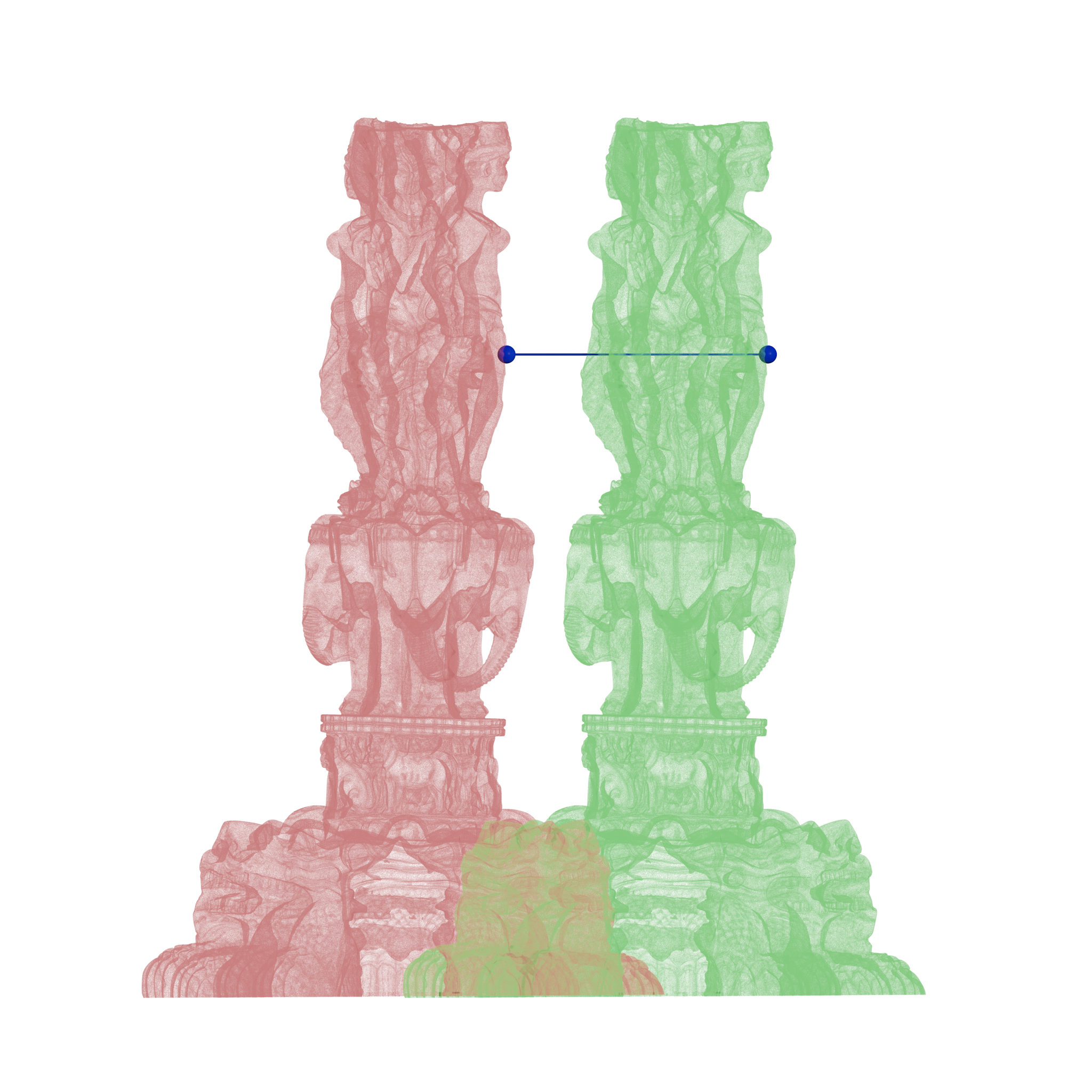}
        \caption{Thai Statue}
    \end{subfigure}
    \begin{subfigure}{0.23\linewidth}
        \includegraphics[width=\textwidth]{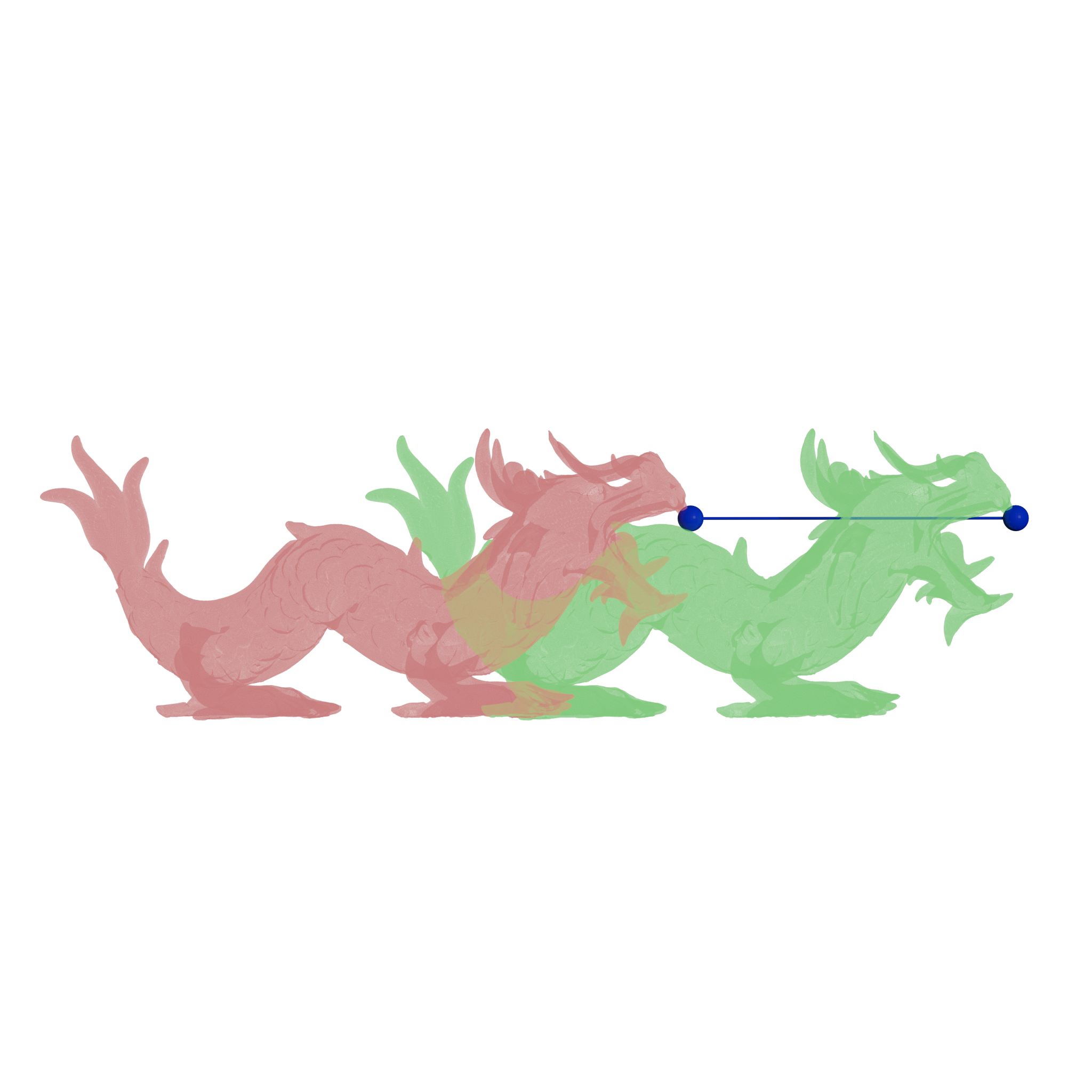}
        \caption{Asian Dragon}
    \end{subfigure}
    
    \caption{Results of the translation benchmarks with a translation ratio of 0.5.}
    \label{fig:translation_result}
\end{figure*}

\begin{figure*}[h]
    \centering
    \begin{subfigure}{0.23\linewidth}
        \includegraphics[width=\textwidth]{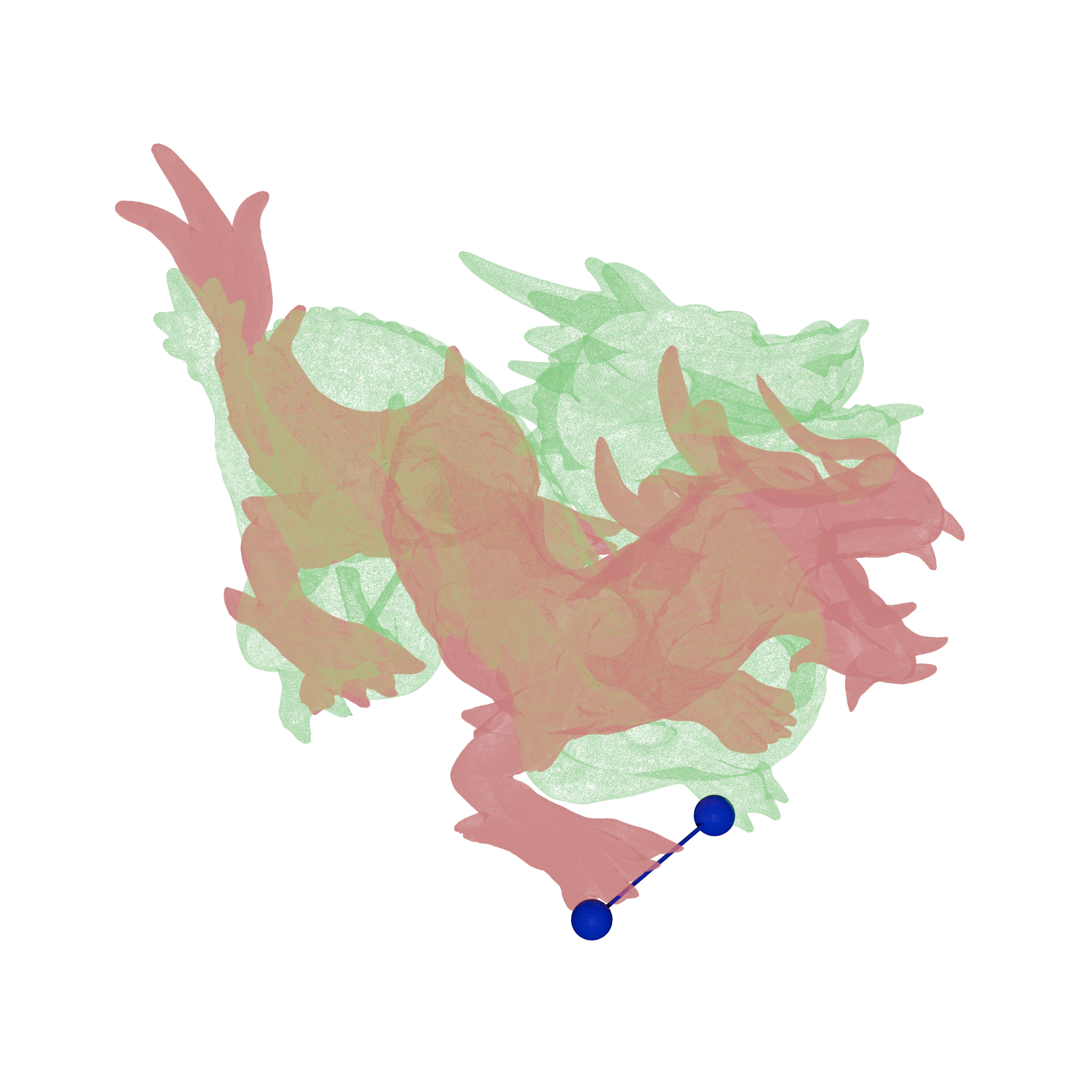}
        \caption{Dragon–Asian Dragon}
    \end{subfigure}
    \begin{subfigure}{0.23\linewidth}
        \includegraphics[width=\textwidth]{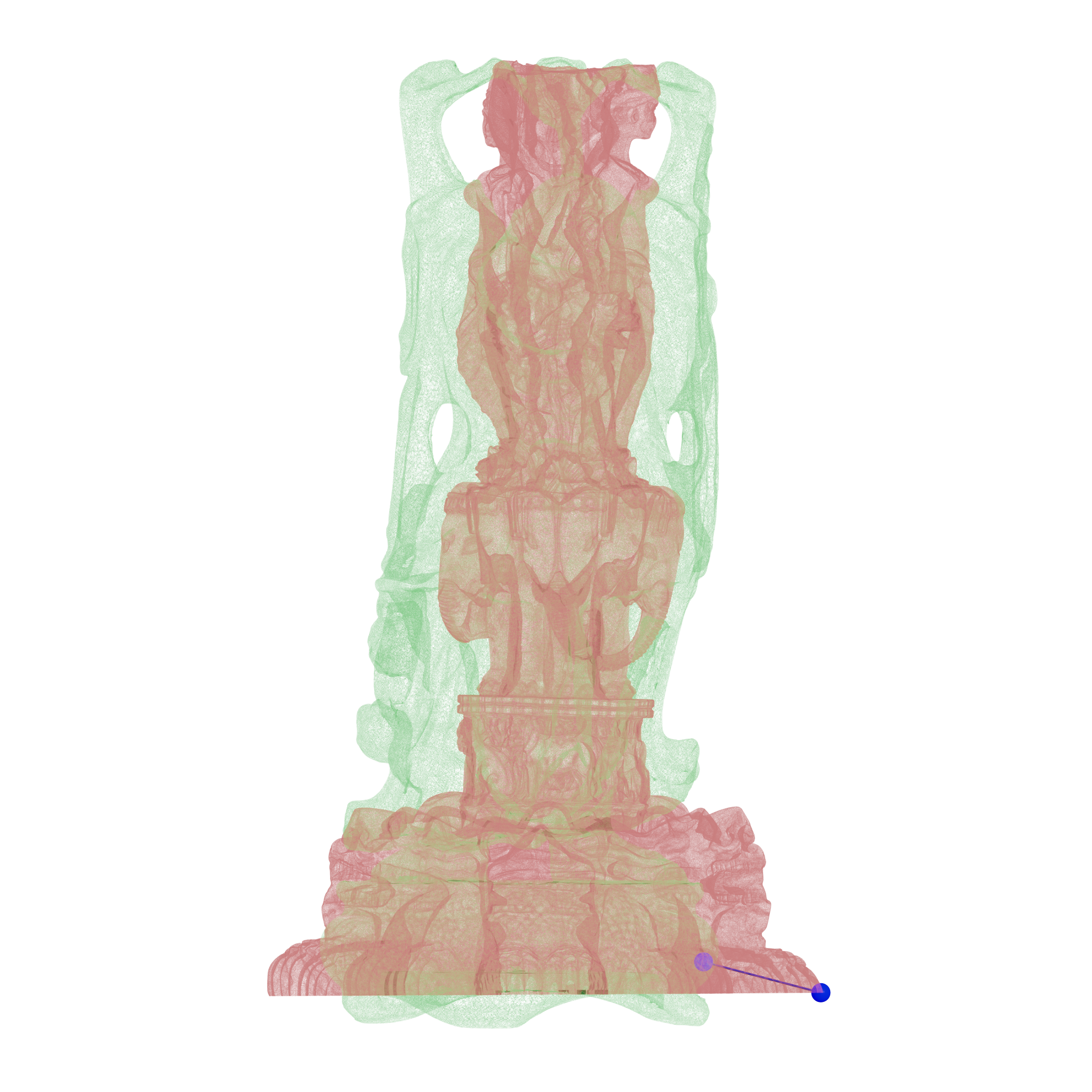}
        \caption{Thai–Buddha}
    \end{subfigure}
    \begin{subfigure}{0.23\linewidth}
        \includegraphics[width=\textwidth]{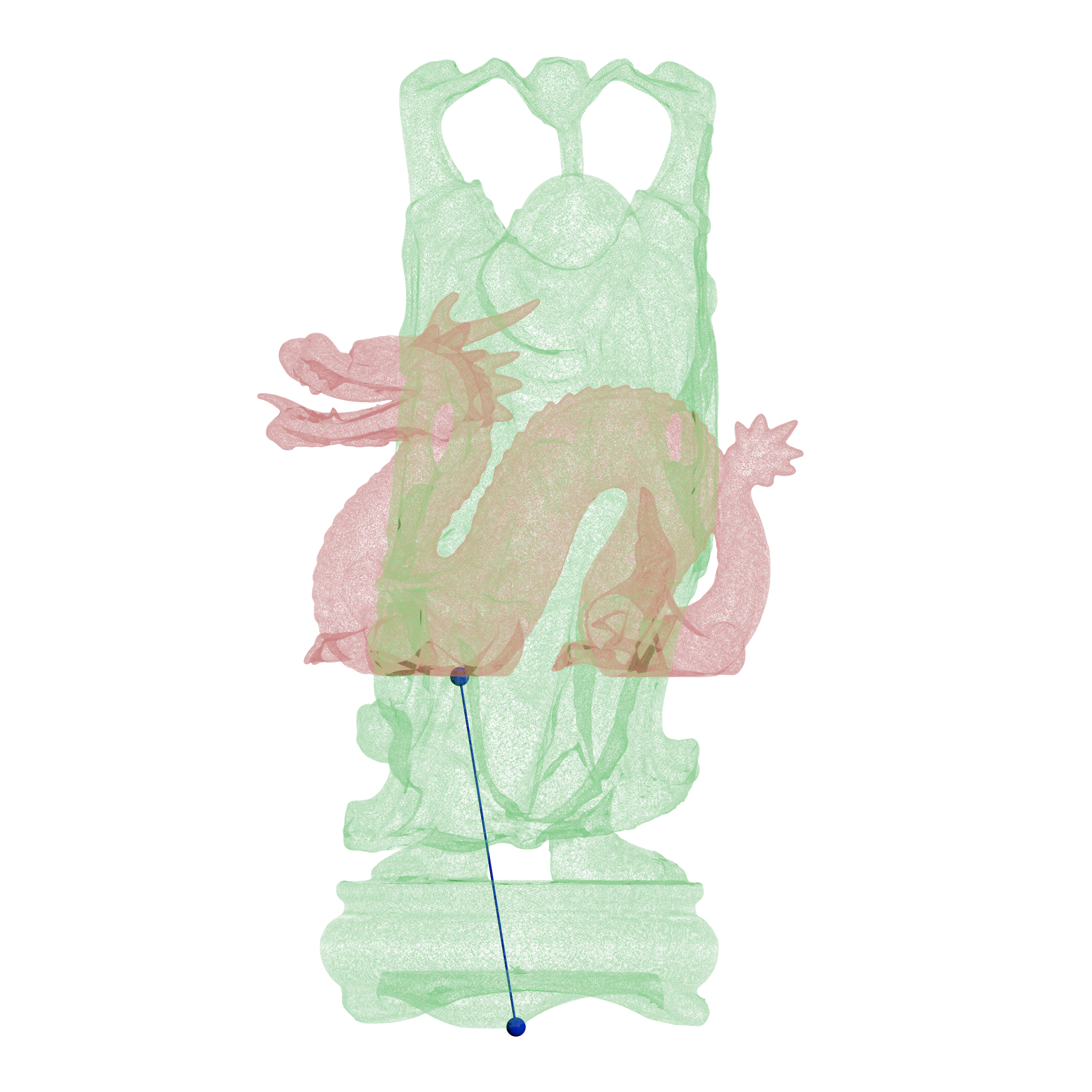}
        \caption{Dragon–Buddha}
    \end{subfigure}
    \begin{subfigure}{0.23\linewidth}
        \includegraphics[width=\textwidth]{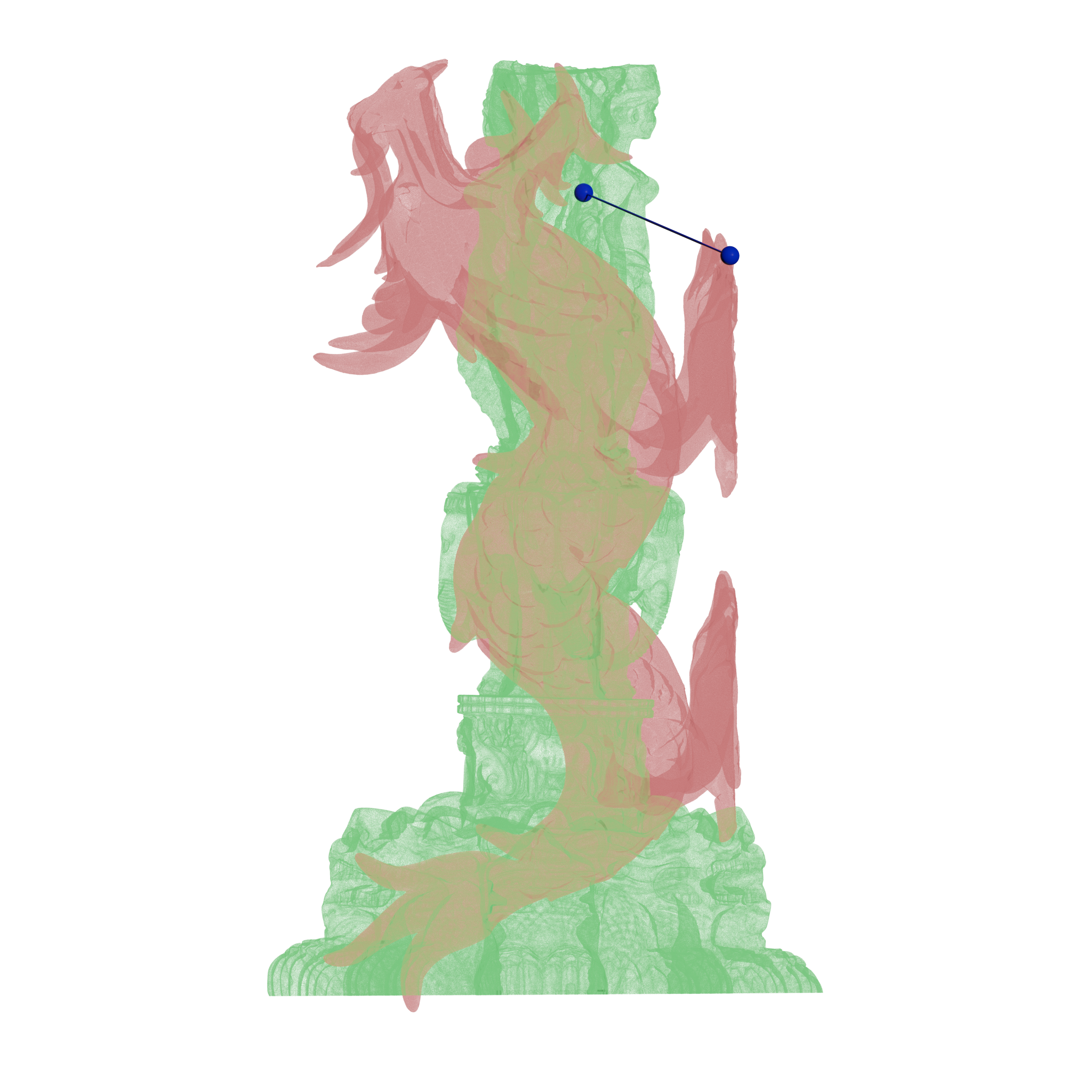}
        \caption{Thai–Asian Dragon}
    \end{subfigure}
    
    \caption{Results of the different object benchmarks.}
    \label{fig:different_object_result}
\end{figure*}

\paragraph{Benchmarks}  
We set up three types of benchmark cases using four well-known models with sizes ranging from 437K to 4.9M vertices (Fig.~\ref{fig:teaser}).
For all experiments, we computed the two-way Hausdorff distance, $H(A, B)$, and validated the results of our method to ensure they matched exactly with those obtained using the brute-force all-pairs comparison method.
\begin{itemize}  
    \item \textbf{Decimation:}  
    This benchmark consists of an original model and its decimated version, both centered at the same location (Fig.~\ref{fig:decimate_result}).
    In this case, the Hausdorff distance serves as a measure of the error introduced by the decimation process.  
    
    \item \textbf{Translation:}  
    This benchmark consists of two identical models, with one of them translated along the an axis by a ratio ranging from 0.3 to 0.7 of the model's length (Fig.~\ref{fig:translation_result}).  
    Here, the Hausdorff distance represents the maximum distance required to align the two models perfectly.  
    
    \item \textbf{Different objects:}  
    This benchmark consists of two different objects placed side by side (Fig.~\ref{fig:different_object_result}).
    In this case, the Hausdorff distance can be used to measure the similarity between the two objects.  
\end{itemize}  

\subsection{Results}

\paragraph{Decimation benchmark}
\begin{table}[t]
\centering
\caption{Comparison of processing times for decimation benchmarks.  
The processing times for our method were measured on the RTX4080.  
For this benchmark, a bit count of 8 was used for Buddha, Dragon, and Statue, and a bit count of 7 was used for Asian Dragon.}  
\label{table:result_decimation}
\small
\begin{tabular}{|c|cc|ccc|}
\hline
\multirow{2}{*}{Model} & \multicolumn{2}{c|}{\# of Vertices}     & \multicolumn{3}{c|}{Processing Time (seconds)}                 \\ \cline{2-6} 
                                & \multicolumn{1}{c|}{Original} & Decimated        & \multicolumn{1}{c|}{\textit{Zheng}} & \multicolumn{1}{c|}{\textit{Sacht}} & \textbf{\textit{Ours}} \\ \hline
Happy Buddha                   & \multicolumn{1}{c|}{543,652}   & 26,986           & \multicolumn{1}{c|}{9.67}           & \multicolumn{1}{c|}{3.69}           & 0.04                   \\ \hline
Dragon                         & \multicolumn{1}{c|}{437,645}   & 22,632           & \multicolumn{1}{c|}{8.92}           & \multicolumn{1}{c|}{3.47}           & 0.04                   \\ \hline
Thai Statue                    & \multicolumn{1}{c|}{4,999,996} & 9,996            & \multicolumn{1}{c|}{27.46}          & \multicolumn{1}{c|}{16.01}          & 0.08                   \\ \hline
Asian Dragon                   & \multicolumn{1}{c|}{3,609,600} & 10,830           & \multicolumn{1}{c|}{21.08}          & \multicolumn{1}{c|}{8.90}           & 0.10                   \\ \hline
\end{tabular}
\end{table}
Table~\ref{table:result_decimation} presents the model sizes and a comparison of processing times across the three algorithms, with our method evaluated on the RTX4080 GPU.
Compared to \textit{Zheng}, \textit{Sacht} demonstrates an average speedup of 2.32x.
However, our RT-HDIST algorithm achieves up to \ToCheck{195.24x} higher performance compared to \textit{Sacht} (\ToCheck{120.97x} on average) and up to \ToCheck{334.87x} higher performance compared to \textit{Zheng}.

Although the performance improvement over \textit{Sacht} slightly decreases when using the RTX2080 and RTX3080, our method still achieves an average speedup of \ToCheck{83.54x} and \ToCheck{89.81x}, respectively.

\paragraph{Translation benchmark}
\begin{table*}[t]
\centering
\caption{Comparison of processing times (in seconds) for translation benchmarks.
The values in parentheses indicate the amount of speed-up than \textit{Zheng}.
For this benchmark, a bit count of 6 was for Dragon, a bit count of 7 was used for Buddha and Asian Dragon, and a bit count of 8 was used for Thai Statue.
}
\label{table:result_trans}
\resizebox{\textwidth}{!}{%
\begin{tabular}{|c|c|c|c|ccc|c|c|c|c|ccc|}
\hline
\multirow{2}{*}{} &
  \multirow{2}{*}{\begin{tabular}[c]{@{}c@{}}Trans.\\ ratio\end{tabular}} &
  \multirow{2}{*}{\textit{Zheng}} &
  \multirow{2}{*}{\textit{Sacht}} &
  \multicolumn{3}{c|}{\textit{\textbf{Ours}}} &
  \multirow{2}{*}{} &
  \multirow{2}{*}{\begin{tabular}[c]{@{}c@{}}Trans.\\ ratio\end{tabular}} &
  \multirow{2}{*}{\textit{Zheng}} &
  \multirow{2}{*}{\textit{Sacht}} &
  \multicolumn{3}{c|}{\textit{\textbf{Ours}}} \\ \cline{5-7} \cline{12-14} 
 &
   &
   &
   &
  \multicolumn{1}{c|}{RTX2080} &
  \multicolumn{1}{c|}{RTX3080} &
  RTX4080 &
   &
   &
   &
   &
  \multicolumn{1}{c|}{RTX2080} &
  \multicolumn{1}{c|}{RTX3080} &
  RTX4080 \\ \hline \hline
\multirow{3}{*}{Dragon} &
   0.3 &
  3.71 &
  2.31 (1.61x) &
  \multicolumn{1}{c|}{0.07 (55.37x)} &
  \multicolumn{1}{c|}{0.07 (57.11x)} &
  0.05 (70.00x) &
\multirow{3}{*}{\begin{tabular}[c]{@{}c@{}}Happy\\ Buddha \end{tabular}} &
  0.3 &
  4.62 &
  2.87 (1.61x) &
  \multicolumn{1}{c|}{0.28 (16.74x)} &
  \multicolumn{1}{c|}{0.20 (23.14x)} &
  0.15 (30.70x) \\ \cline{2-7} \cline{9-14} 
 &
  0.5 &
  3.72 &
  2.42 (1.54x) &
  \multicolumn{1}{c|}{0.08 (44.66x)} &
  \multicolumn{1}{c|}{0.07 (50.22x)} &
  0.06 (59.34x) &
   &
  0.5 &
  4.53 &
  2.88 (1.57x) &
  \multicolumn{1}{c|}{0.26 (17.37x)} &
  \multicolumn{1}{c|}{0.18 (25.65x)} &
  0.13 (34.70x) \\ \cline{2-7} \cline{9-14} 
 &
  0.7 &
  3.69 &
  2.27 (1.63x) &
  \multicolumn{1}{c|}{0.09 (42.26x)} &
  \multicolumn{1}{c|}{0.08 (46.89x)} &
  0.06 (60.70x) &
   &
  0.7 &
  4.61 &
  2.87 (1.60x) &
  \multicolumn{1}{c|}{0.32 (14.59x)} &
  \multicolumn{1}{c|}{0.21 (22.41x)} &
  0.15 (30.72x)
  \\ \hline \hline
\multirow{3}{*}{\begin{tabular}[c]{@{}c@{}}Thai\\ Statue\end{tabular}} &
  0.3 &
  50.28 &
  41.32 (1.22x) &
  \multicolumn{1}{c|}{1.58 (31.81x)} &
  \multicolumn{1}{c|}{1.08 (46.49x)} &
  0.78 (64.86x) &
  \multirow{3}{*}{\begin{tabular}[c]{@{}c@{}}Asian\\  Dragon\end{tabular}} &
  0.3 &
  33.41 &
  25.33 (1.32x) &
  \multicolumn{1}{c|}{0.37 (89.78x)} &
  \multicolumn{1}{c|}{0.33 (101.42x)} &
  0.25 (134.95x) \\ \cline{2-7} \cline{9-14}
 &
  0.5 &
  50.13 &
  43.18 (1.16x) &
  \multicolumn{1}{c|}{2.20 (22.74x)} &
  \multicolumn{1}{c|}{1.39 (36.00x)} &
  1.00 (50.14x) &
   &
  0.5 &
  33.36 &
  25.10 (1.33x) &
  \multicolumn{1}{c|}{0.23 (143.07x)} &
  \multicolumn{1}{c|}{0.20 (167.24x)} &
  0.14 (232.68x) \\ \cline{2-7} \cline{9-14} 
 &
  0.7 &
  50.37 &
  43.56 (1.16x) &
  \multicolumn{1}{c|}{4.13 (12.19x)} &
  \multicolumn{1}{c|}{2.46 (20.48x)} &
  1.76 (28.56x) &
   &
  0.7 &
  33.38 &
  24.09 (1.39x) &
  \multicolumn{1}{c|}{0.20 (169.98x)} &
  \multicolumn{1}{c|}{0.17 (194.41x)} &
  0.13 (250.37x)
 \\ \hline
\end{tabular}%
}
\end{table*}

Table~\ref{table:result_trans} presents the processing times of the three algorithms across different scene setups.
Overall, \textit{Sacht} achieved up to \ToCheck{1.63x} (\ToCheck{1.43x} on average) higher performance compared to \textit{Zheng}.
In contrast, our method, running on the RTX4080, demonstrated up to \ToCheck{250.37x} and \ToCheck{180.72x} (\ToCheck{87.31x} and \ToCheck{63.24x} on average, respectively) higher performance compared to \textit{Zheng} and \textit{Sacht}.

With the RTX2080 and RTX3080, our method demonstrates \ToCheck{55.05x} and \ToCheck{65.95x} higher performance on average compared to \textit{Zheng}, respectively.
These results also indicate that our RT-HDIST achieves better performance with higher-performance GPUs.

The performance improvement of our method over alternatives varies depending on the scene setup (e.g., models and their placement).
Different scene setups create varying point distributions, which can result in longer iterations during the RT-accelerated candidate search step or larger final candidate sets.
We observed that both cases lead to reduced performance improvement.
Additionally, while we found that the best performance could be achieved with different bit counts for each scene setup, we used a fixed bit count per model to report practical and consistent performance results.

Nonetheless, our RT-HDIST consistently outperformed prior methods in all cases, achieving significant performance improvements in most scenarios.
It is also worth noting that the performance multiplier is highly sensitive, as our method's processing time operates at the millisecond scale in most cases, whereas alternative methods take several seconds or even tens of seconds.

\paragraph{Different object benchmark}
\begin{table}[t]
\centering
\caption{Comparison of processing times (in seconds) for different objects benchmark.
$k$ is the bit count used for each case.
}
\label{table_result_diff}
\resizebox{\columnwidth}{!}{%
\begin{tabular}{|c|c|c|c|ccc|}
\hline
\multirow{2}{*}{Models} & \multirow{2}{*}{$k$} & \multirow{2}{*}{\textit{Zheng}} & \multirow{2}{*}{\textit{Sacht}} & \multicolumn{3}{c|}{\textit{\textbf{Ours}}} \\ \cline{5-7} 
                  &   &       &       & \multicolumn{1}{c|}{RTX2080} & \multicolumn{1}{c|}{RTX3080} & RTX4080 \\ \hline
Dragon-Asian Dra. & 7 & 18.68 & 11.57 & \multicolumn{1}{c|}{0.30}    & \multicolumn{1}{c|}{0.25}    & 0.18    \\ \hline
Thai-Buddha       & 7 & 26.14 & 16.86 & \multicolumn{1}{c|}{0.57}    & \multicolumn{1}{c|}{0.44}    & 0.31    \\ \hline
Dragon-Buddha     & 6 & 4.11  & 2.57  & \multicolumn{1}{c|}{0.20}    & \multicolumn{1}{c|}{0.18}    & 0.14    \\ \hline
Thai-Asian Dra.   & 7 & 41.60 & 30.51 & \multicolumn{1}{c|}{0.95}    & \multicolumn{1}{c|}{0.76}    & 0.57    \\ \hline
\end{tabular}%
}
\end{table}

\Skip{
\begin{table}[t]
\centering
\caption{Comparison of processing times (in seconds) for different objects benchmark.}
\label{table_result_diff}
\resizebox{\columnwidth}{!}{%
\begin{tabular}{|cc|c|c|ccc|}
\hline
\multicolumn{2}{|c|}{Objects} & \multirow{2}{*}{\textit{Zheng}} & \multirow{2}{*}{\textit{Sacht}} & \multicolumn{3}{c|}{\textit{\textbf{Ours}}} \\ \cline{1-2} \cline{5-7} 
\multicolumn{1}{|c|}{A}      & B          &  &  & \multicolumn{1}{c|}{RTX2080} & \multicolumn{1}{c|}{RTX3080} & RTX4080 \\ \hline
\multicolumn{1}{|c|}{Dragon} & Asian Dra. 
& 18.68
& 11.57 (1.62x)
& \multicolumn{1}{c|}{0.30 (63.00x)}
& \multicolumn{1}{c|}{0.25 (75.73x)}
&  0.18 (102.93x)       \\ \hline
\multicolumn{1}{|c|}{Thai}   & Buddha     
& 26.14
& 16.86 (1.55x)
& \multicolumn{1}{c|}{0.57 (46.15x)}       
& \multicolumn{1}{c|}{0.44 (59.42x)}       
& 0.31 (83.80x)        \\ \hline
\multicolumn{1}{|c|}{Dragon} & Buddha     
& 4.11
& 2.57 (1.60x)
& \multicolumn{1}{c|}{0.20 (20.49x)}       
& \multicolumn{1}{c|}{0.18 (23.00x)}       
& 0.14 (29.51x)        \\ \hline
\multicolumn{1}{|c|}{Thai}   & Asian Dra. 
&  41.60
&  30.51 (1.36x)
& \multicolumn{1}{c|}{0.95 (43.93x)}       
& \multicolumn{1}{c|}{0.76 (54.93x)}       
&  0.57 (73.07x)       \\ \hline
\end{tabular}%
}
\end{table}
}
Table~\ref{table_result_diff} presents the processing times of the three algorithms across different combinations of two objects.
Similar to the other benchmarks, our method on the RTX4080 achieved up to \ToCheck{140.44x} and \ToCheck{75.53x} (\ToCheck{56.98x} and \ToCheck{33.39x} on average) higher performance compared to \textit{Zheng} and \textit{Sacht}, respectively.
These results demonstrate that our method consistently outperforms prior methods across various scenarios.

\subsection{Performance Analysis}
\begin{figure}[t]
    \centering
    \includegraphics[width=\linewidth]{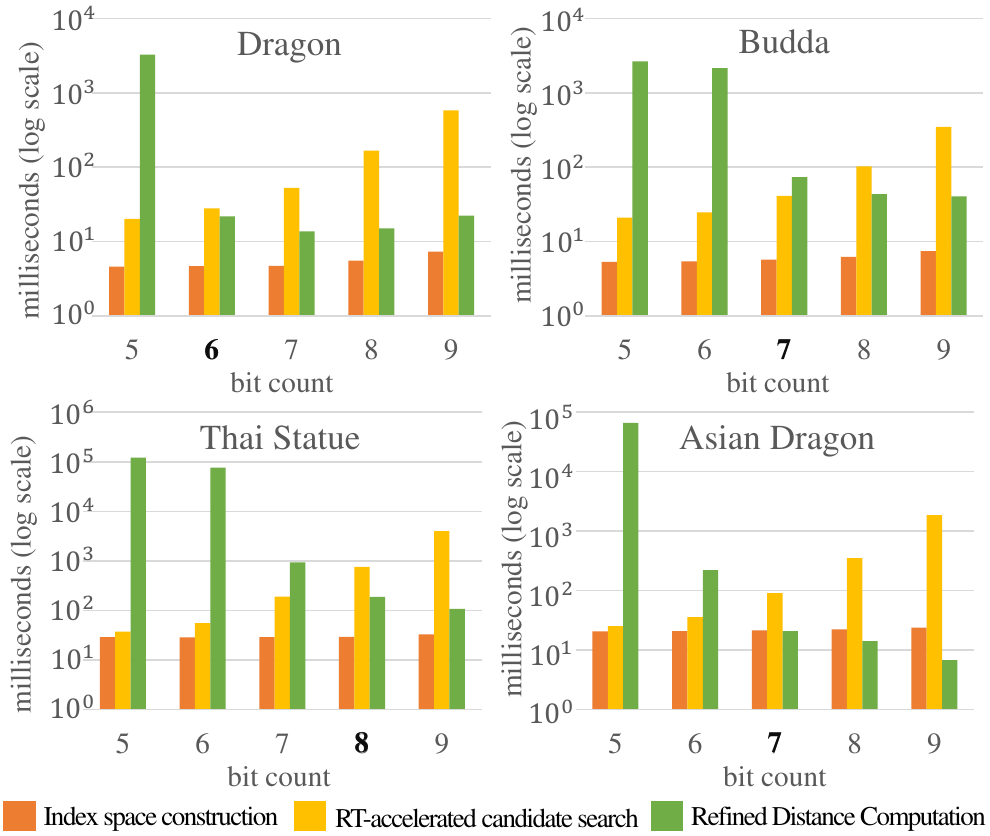}
    \caption{Processing time (log-scale) of each component of our method on RTX4080.  
    For this analysis, we used the translation benchmark with a ratio of 0.5.  
    The bold font in the bit count indicates the best performance case.  
    Similar trends were observed for other benchmarks and GPUs.}  
    \label{fig:analysis}
\end{figure}  

Our RT-HDIST algorithm consists of three main components, and the contribution of each component to the total processing time varies depending on the benchmark scenes and the bit count (i.e., $k$, a hyperparameter of our method).  
To analyze how the bit count impacts the processing time of each component and to observe performance trends, we measured and evaluated the processing time for each step individually across different bit counts.  
Fig.~\ref{fig:analysis} presents the results for the translation benchmark.

The index space construction step exhibits relatively stable processing times across different bit counts, with a slight increase observed for larger bit counts.  
This step accounts for approximately \ToCheck{0.1}–\ToCheck{7.5}\% of the total processing time.  

On the other hand, the second step (RT-accelerated candidate search) and the third step (refined distance computation) show an inverse relationship in their processing time contributions.  
As the bit count increases (i.e., the cell size becomes smaller), the search radius increment decreases.  
This results in an increased number of iterations during the RT-accelerated candidate search step, while simultaneously reducing the number of final candidates for the refined computation step.  
Consequently, as the bit count increases, the processing time for the second step increases, whereas the processing time for the third step decreases.  

Generally, our method achieves optimal performance when the two major steps—RT-accelerated candidate search and refined distance computation—are balanced in their processing time contributions (i.e., have similar portions).  
For example, in the best-performing cases shown in Fig.~\ref{fig:analysis}, the RT-accelerated candidate search and refined distance computation account for \ToCheck{57.83}\% and \ToCheck{34.03}\% of the total processing time on average, respectively. 
This balance is typically observed with bit counts ranging from 6 to 8, and larger-scale scenes generally require higher bit counts.  

\subsection{Effect Of Using Index Space}
\begin{table}[t]
\centering
\caption{Comparison of processing times (in seconds) between \textit{Ours-NoIS} and \textit{Ours} for the different objects benchmark.  
Values in parentheses represent the amount of processing time reduction.  }
\label{table_result_OursNoIS_vs_Ours}
\resizebox{\columnwidth}{!}{%
\begin{tabular}{|c|c|c|c|c|}
\hline
                   & Dragon-Asian D. & Tahi-Buddha   & Dragon-Buddha & Thai-Asian D. \\ \hline
\textit{Ours-NoIS} & 1.99                & 1.19          & 1.29          & 37.00           \\ \hline
\textit{Ours}      & 0.18 (90.90\%)       & 0.31 (73.84\%) & 0.14 (89.18\%) & 0.57 (98.46\%)   \\ \hline
\end{tabular}%
}
\end{table}

To evaluate the benefits of using index space compared to performing RT-accelerated candidate search directly in the object space, we implemented an alternative version of our method without index space construction (i.e., \textit{Ours-NoIS}).  
For \textit{Ours-NoIS}, we set the same initial radius and increment as in the case of \textit{Ours}, specifically $cell\_size * \sqrt{3}$.  

Table~\ref{table_result_OursNoIS_vs_Ours} compares the performance of \textit{Ours} and \textit{Ours-NoIS} for the different object benchmark, evaluated on the RTX4080.  
We observed that the RT-accelerated candidate search step takes significantly more time in \textit{Ours-NoIS} compared to \textit{Ours}, as the number of points to process in each iteration is considerably larger without the clustering provided by index space.  

Moreover, the performance gap widens in scenarios requiring more iterations in the RT-accelerated candidate search step, such as when the distance between the two objects increases.  
Overall, with index space construction, our method reduces the processing time by \ToCheck{70.89}\% on average across all benchmarks used in our experiments.  

These results validate the significant advantages of using index space in our method, demonstrating its effectiveness in reducing computational overhead and improving performance.


\section{Conclusion}  
In this work, we introduced RT-HDIST, the first RT-core-based algorithm for Hausdorff distance computation.  
By reformulating the Hausdorff distance problem as a series of nearest-neighbor searches and leveraging RT-cores with index space optimization, we achieved significant performance improvements while maintaining exact results.  
Extensive benchmarks demonstrated that RT-HDIST consistently outperforms state-of-the-art methods, achieving up to two orders of magnitude speedup across various scenarios.  
These results highlight RT-HDIST's potential for real-time and large-scale applications in computer graphics, robotics, and beyond.  

There are two hyperparameters in our method: bit count ($k$) and initial radius ($r_0$).  
While a bit count in the range of 6 to 8 and using the center distance between two objects for $r_0$ generally yield good performance, we observed that fine-tuning these parameters can further improve performance depending on the specific scenario.  
As future work, we aim to develop an algorithm that can automatically select the optimal hyperparameters to achieve the best performance for varying use cases.

\bibliographystyle{abbrv-doi}
\bibliography{ref}

\newpage

\end{document}